\documentclass[prd,twocolumn,amsmath,amssymb]{revtex4}

\def\al{\alpha}
\def\be{\beta}
\def\ga{\gamma}
\def\de{\delta}
\def\ep{\epsilon}

\def\et{\eta}
\def\th{\theta}

\def\ka{\kappa}
\def\la{\lambda}

\def\rh{\rho}

\def\si{\sigma}

\def\ta{\tau}

\def\ph{\phi}

\def\ch{\chi}
\def\ps{\psi}

\def\La{\Lambda}
\def\Si{\Sigma}

\def\Ph{\Phi}

\def\cl{{\cal L}}

\def\cF{{\cal F}}
\def\cl{{\cal L}}
\def\cL{{\cal L}}

\def\fr#1#2{{{#1} \over {#2}}}
\def\frac#1#2{{\textstyle{{#1}\over {#2}}}}
\def\half{{\textstyle{1\over 2}}}

\def\vev#1{\langle {#1}\rangle}

\def\prt{\partial}

\def\lsim{\mathrel{\rlap{\lower4pt\hbox{\hskip1pt$\sim$}}
    \raise1pt\hbox{$<$}}}
\def\gsim{\mathrel{\rlap{\lower4pt\hbox{\hskip1pt$\sim$}}
    \raise1pt\hbox{$>$}}}
\def\sqr#1#2{{\vcenter{\vbox{\hrule height.#2pt
         \hbox{\vrule width.#2pt height#1pt \kern#1pt
         \vrule width.#2pt}
         \hrule height.#2pt}}}}

\def\pt#1{\phantom{#1}}
\def\ol#1{\overline{#1}}

\def\twiddle{\lower4pt\hbox{\hskip0pt{$\widetilde{}$}}}
\def\m@th{\mathsurround=0pt}
\def\cmapstochar{\mathrel{\rlap{
  \lower0.1pt\hbox{\hskip0.25pt{$\mapstochar$}}}
  \raise0pt\hbox{\hskip4.5pt{$\twiddle$}}}}
\def\notsimfill{$\m@th\cmapstochar$}
\def\scroodle#1{\vbox{\ialign{##\crcr\notsimfill\crcr
  \noalign{\kern-4pt\nointerlineskip}
   $\hfil\displaystyle{#1}\hfil$\crcr}}}

\newcommand{\beq}{\begin{equation}}
\newcommand{\eeq}{\end{equation}}
\newcommand{\bea}{\begin{eqnarray}}
\newcommand{\eea}{\end{eqnarray}}
\newcommand{\rf}[1]{(\ref{#1})}

\def\etal{{\it et al.}}

\def\kl{{\ka\la}}
\def\km{{\ka\mu}}
\def\kn{{\ka\nu}}
\def\lm{{\la\mu}}
\def\ln{{\la\nu}}
\def\mn{{\mu\nu}}
\def\lmn{{\la\mu\nu}}
\def\klmn{{\ka\la\mu\nu}}

\def\cb{{\mathfrak b}}
\def\cB{{\mathfrak B}}
\def\cH{{\mathfrak H}}

\def\sb{\overline{s}}
\def\tb{\overline{t}}
\def\ub{\overline{u}}

\def\Btw{\scroodle{B}}

\def\nsy{\et_{\mu\nu}}

\def\Vxxb{\la_{11}}
\def\Vxyb{\la_{12}}
\def\Vyyb{\la_{22}}

\def\Xxtw{X_1}
\def\Xytw{X_2}

\def\Xop{X_1^\prime}
\def\Xtp{X_2^\prime}

\begin{document}

\title{Lorentz violation with an antisymmetric tensor} 

\author{
Brett Altschul,$^1$ 
Quentin G.\ Bailey,$^2$
and V.\ Alan Kosteleck\'y$^3$
}

\affiliation{
$^1$Department of Physics and Astronomy,
University of South Carolina,
Columbia, South Carolina 29208, USA\\
$^2$Physics Department, Embry-Riddle Aeronautical University,
Prescott, Arizona 86301, USA\\
$^3$Physics Department, Indiana University, 
Bloomington, Indiana 47405, USA
}

\date{IUHET 537, December 2009} 

\begin{abstract}
Field theories with spontaneous Lorentz violation 
involving an antisymmetric 2-tensor are studied.
A general action including nonminimal gravitational couplings
is constructed,
and features of the Nambu-Goldstone and massive modes
are discussed.
Minimal models in Minkowski spacetime exhibit 
dualities with Lorentz-violating vector and scalar theories.
The post-newtonian expansion for nonminimal models 
in Riemann spacetime
involves qualitatively new features,
including the absence of an isotropic limit.
Certain interactions producing stable Lorentz-violating theories
in Minkowski spacetime 
solve the renormalization-group equations 
in the tadpole approximation.
\end{abstract}


\maketitle

\section{Introduction}

Among the simpler field theories in Minkowski spacetime
are ones built from $p$-forms.
These include electrodynamics, 
which is the abelian gauge theory of a 1-form,
and field theories constructed with antisymmetric $p$-tensors.
The predominant examples of the latter
include models with a gauge-invariant kinetic term
for an antisymmetric 2-tensor,
sometimes called the notoph 
\cite{op}
or the Kalb-Ramond field
\cite{mkpr}. 
These theories have some elegant properties,
including dualities to other $p$-form theories
\cite{op,mkpr,csnkht}.

In this work,
we consider Lorentz-violating field theories 
with an antisymmetric 2-tensor,
including models coupled to gravity.
In a generic Lagrange density,
terms can be constructed 
that explicitly violate Lorentz symmetry
by forming observer invariants
from tensor operators and $c$-number coefficients. 
However,
explicit breaking is generically incompatible
with the Bianchi identities in Riemann geometry
and hence is problematic for theories with gravity 
\cite{akgrav}.
A viable alternative is spontaneous Lorentz violation,
in which a potential term 
drives the development of a nonzero vacuum value
for a tensor field 
\cite{ksp}.
In theories of this type,
the Lagrange density is Lorentz invariant,
but the presence of the tensor vacuum value 
means the physics can display Lorentz breaking.
Here,
our focus is on theories with spontaneous Lorentz violation
triggered by a potential for an antisymmetric 2-tensor field.

Spontaneous Lorentz violation
triggered by a potential for an arbitrary tensor field 
is accompanied by certain generic features.
Massless Nambu-Goldstone (NG) modes
\cite{ng}
emerge that are 
associated with field fluctuations
along the broken Lorentz generators
\cite{bk}.
When a smooth potential drives the Lorentz breaking,
massive modes can also appear
\cite{bfk}.
The role of the NG and massive modes
is central to the physical content
of a field theory with spontaneous Lorentz violation.
Some of their properties are generic to any field theory,
while others depend on the specific field content
and structure of the action.
One goal here is to apply this work
to theories based on an antisymmetric 2-tensor field,
establishing some basic properties
of the corresponding NG and massive modes.

Another motivation for this work stems from the possibility
of novel experimental signals for Lorentz violation
involving gravitational couplings.
Recent years have seen extensive tests 
of Lorentz symmetry in Minkowski spacetime
\cite{datatables},
but the scope of searches involving gravitational couplings
remains comparatively limited.
Dominant curvature couplings involving Lorentz violation
are controlled by three coefficient fields,
conventionally denoted as
$s^\mn$, $t^\klmn$, and $u$
\cite{akgrav}. 
Constraints on some $s^\mn$ coefficients 
have been attained using lunar laser ranging
\cite{bcs}
and atom interferometry
\cite{mchcc},
and numerous other experimental and observational signals
from these couplings can arise at the post-newtonian level
\cite{qbak}.
However,
to date no gravitational field theory 
has been constructed to yield nonzero $t^\klmn$ coefficients.
We show here that a comparatively simple coupling of this type
can appear in theories involving 
gravitational couplings to an antisymmetric 2-tensor.
The post-newtonian expansion is affected
in a purely anisotropic way,
offering a qualitatively distinct source
of signals for Lorentz violation.

In Minkowski spacetime,
spontaneous Lorentz violation arises
whenever the potential for the interactions
has a nontrivial stable extremum.
An interesting issue is the behavior 
of the Lorentz-violating interactions
under quantum corrections.
In certain vector models with spontaneous Lorentz violation,
nontrivial potentials solve 
the renormalization-group (RG) equations
in the tadpole approximation
\cite{baak}.
Part of the present work revisits
this possibility in the context
of a minimal theory with an antisymmetric 2-tensor.
We investigate the RG flow in the tadpole approximation 
and obtain analytical solutions for the potential.
All the nontrivial stable potentials that result
describe theories with spontaneous Lorentz violation. 

The organization of this work is as follows.
We begin in Sec.\ \ref{basics}
with the basic construction of the field theory,
including the gravitational couplings.
Some general features of the potential
and consequences of the spontaneous Lorentz breaking
are discussed.
In Sec.\ \ref{Minimal model},
properties of minimal models in Minkowski spacetime
are established,
including correspondences to dual theories. 
Sec.\ \ref{Nonminimal model}
focuses on gravitational couplings
and their consequences for post-newtonian physics.
The field equations are obtained,
linearization is performed in a Minkowski background,
and the post-newtonian metric is obtained
at third order.
In Sec.\ \ref{RGanalysis},
we return to the minimal model in Minkowski spacetime
and implement the tadpole approximation 
for the RG equations,
obtaining solutions for the potential.
A summary is provided in Sec.\ \ref{Summary}.
Throughout this work,
we adopt the conventions of Ref.\ \cite{akgrav}.

\section{Field theory}
\label{basics}

In this section,
the action for an antisymmetric 2-tensor 
in four-dimensional Riemann spacetime is considered.
Some definitions are introduced,
and basic properties of the field theory are summarized.
We construct the nonminimal gravitational couplings
and discuss some general features associated with the potential 
driving spontaneous Lorentz violation.

\subsection{Setup}

The fundamental field of interest in this work
is an antisymmetric 2-tensor,
denoted $B_{\mu\nu} = -B_{\nu\mu}$. 
It is convenient to introduce the dual tensor $\cB_\mn$,
defined by
\beq
\cB_\mn \equiv \half \ep_{\mu\nu\ka\la} B^{\ka\la},
\label{dual}
\eeq
where $\ep_\klmn$ 
is the totally antisymmetric Levi-Civita tensor. 
The covariant derivative of $B_\mn$ is denoted $D_\la B_\mn$.
This work considers Riemann spacetimes,
and the covariant derivative 
is constructed with the Levi-Civita connection.
The generalization to the Cartan connection 
and the corresponding Riemann-Cartan spacetimes with torsion
\cite{torsion} 
is of potential interest,
although current experimental constraints 
\cite{torexp}
suggest nonzero torsion components
are likely to have at most a limited phenomenological impact.

A useful combination of derivatives 
is the totally antisymmetric field-strength tensor $H_\lmn$,
given by
\bea
H_\lmn &=& 
\prt_\la B_{\mu\nu} + \prt_\mu B_{\nu\la} + \prt_\nu B_{\la\mu}
\label{h}                 
\eea
and its dual 
\beq
\cH_\ka \equiv \frac 1 6 \ep_{\klmn} H^{\lmn}.
\label{hdual}
\eeq
The field strength $H_\lmn$
can be viewed as the components of an exact 3-form field $H$
constructed via the exterior derivative
from the 2-form $B$ associated with $B_\mn$.
Covariant derivatives can also be used in Eq.\ \rf{h} 
because the connection coefficients cancel
in Riemann spacetime.
The field strength $H_\lmn$ satisfies the identity
\beq
\prt_\ka H_{\la\mu\nu} -\prt_\la H_{\mu\nu\ka}
+\prt_\mu H_{\nu\ka\la} -\prt_\nu H_{\ka\la\mu} = 0 ,
\label{cyclic}
\eeq
which follows because an exact 3-form is closed.
Again,
covariant derivatives can be used in this expression instead.

The field strength $H_\lmn$ is invariant 
under a gauge transformation of $B_\mn$ given by
\beq
B_\mn \to B_\mn + \prt_\mu \La_\nu - \prt_\nu \La_\mu ,
\label{gaugetr}
\eeq
which represents a shift of $B$ by an exact 2-form.
The gauge parameter $\La_\mu$ has four components,
but the transformation involves only three independent effects
because the shift 
\beq
\La_\mu \to \La_\mu + \prt_\mu \Si
\eeq
leaves Eq.\ \rf{gaugetr} unchanged.
This latter shift represents
a subsidiary gauge transformation
involving an exact 1-form.

The action for the theory including
gravitational and matter sectors can be written as
\beq
S=\int d^4 x ~ e (\cl_g + \cl_M + \cl_B + \cl_V ) ,
\label{totact}
\eeq
where $e$ is the metric determinant
and the Lagrange density $e\cL$
is split into four pieces,
corresponding to the pure-gravity sector
$e\cl_g$,
the matter sector $e\cl_M$,
the $B_\mn$ kinetic term $e\cl_B$,
and the potential term $e\cl_V$.
For our purposes,
it suffices for $e\cl_g$ to adopt 
the usual Einstein-Hilbert action of general
relativity with cosmological constant $\La$,
\beq
e\cl_g = \fr e{2\ka} (R-2\La),
\label{eh}
\eeq
where $\ka=8\pi G_N$
with $G_N$ the Newton gravitational constant.
Also,
the specific content of the matter-sector Lagrange density $e\cl_M$ 
is secondary here,
and in much of the analysis to follow
it suffices to assume vanishing matter couplings to $B_\mn$.
When useful,
a matter coupling to $B_\mn$ can be introduced
in the form
\beq
e\cl_M \supset -\half e B_\mn j_B^\mn,
\eeq
where $j_B^\mn$ is the corresponding current.
This coupling is analogous 
to that of the Kalb-Ramond current in string theory
\cite{mkpr}.
For certain actions of the form \rf{totact},
including ones in Minkowski spacetime that are 
invariant under the gauge transformation \rf{gaugetr},
the current $j_B^\mn$ is conserved. 
Note also that a nonzero vacuum value for $B_\mn$ 
can lead to terms in the effective action
of the type found in the minimal
Standard-Model Extension (SME)
\cite{ck}.
For example,
a current $j_B ^\mn = \ol{\ps}\si^\mn\ps$
generates an SME coefficient of the $H_\mn$ type.

\subsection{Kinetic term}
\label{Kinetic term}

By definition,
the kinetic term $e\cl_B$ in the action \rf{totact}
determines the dynamics of $B_\mn$,
including its nonminimal couplings to gravity.
In this work,
we restrict attention 
to kinetic terms of second order in derivatives of $B_\mn$. 
For some of the analysis,
it is useful to allow also
nonminimal nonderivative gravitational couplings
that are linear in the curvature tensor.
Higher-order derivative couplings 
associated with Lorentz violation
could in principle be incorporated,
at least at the level of effective field theory
\cite{kp}.
A classification of all derivative operators
might be achieved following the methodology 
adopted for Lorentz-violating electrodynamics
\cite{km},
but this lies beyond our present scope.

In the present subsection,
we provide the general Lagrange density $\cl_B$
containing all independent quadratic kinetic terms 
for the antisymmetric tensor $B_\mn$,
along with all independent nonminimal nonderivative couplings 
to gravity that are linear in the curvature.
It is convenient to split the Lagrange density $\cl_B$
into two parts,
\beq
e\cl_B=e\cl_{BB} + e\cl_{B\cB},
\label{Bact}
\eeq
where the parity-even term $\cl_{BB}$ 
involves quadratic expressions in $B_\mn$
and the parity-odd term $\cl_{B\cB}$
involves the product of $B_\mn$ and $\cB_\mn$. 

The general form of the parity-even term 
can be written as 
\bea
e\cl_{BB} &=&  
\ta_1 eH_{\lmn} H^{\lmn} 
+\ta_2 e
(D_\la B^\ln) 
(D_\mu B^\mu_{\pt{\la}\nu})
\nonumber\\
&&
+\ta_3 eB^{\kl} B^{\mn} R_{\klmn} 
+\ta_4 eB^{\ln} B^\mu_{\pt{\la}\nu} R_{\lm}
\nonumber\\
&&
+\ta_5 eB^\mn B_\mn R ,
\label{sbb}
\eea
where 
$\ta_1$,
$\ta_2$,
$\ta_3$,
$\ta_4$,
$\ta_5$
are arbitrary constants.
Note that only the first term in this expression
is invariant under the gauge transformation \rf{gaugetr}. 
In constructing Eq.\ \rf{sbb},
we can omit the two scalars
$(D_\la B_\mn) (D^\mu B^{\nu\la})$ 
and 
$(D_\la B^\mn) (D^\la B_\mn)$ 
because they
are equivalent to other terms 
via the identity 
\beq
H_{\lmn} H^{\lmn}= 
3 (D_\la B^\mn) (D^\la B_\mn)
+ 6 (D_\la B_\mn) (D^\mu B^{\nu\la})
\label{Hsq}
\eeq
and the integral relation 
\bea
\int d^4x ~e 
\Big[ 
(D_\la B_\mn) (D^\mu B^{\nu\la})
+(D_\la B^\ln)(D_\mu B^{\mu}_{\pt{\la}\nu})
\nonumber\\
- B^{\ln} B^\mu_{\pt{\la}\nu} R_{\lm}
+\half B^{\kl} B^{\mn} R_{\klmn} \Big] = 0.
\label{intrel}
\eea
This last relation holds up to surface terms,
which leave unaffected the equations of motion.

The general form of the parity-odd term
involving both $B_\mn$ and its dual $\cB_\mn$
can be written as:
\bea
e\cl_{B\cB} &=&    
\si_1 e (D_\la B_\mn) (D^\la \cB^\mn) 
+\si_2 e (D_\la B^{\la}_{\pt{\la}\nu}) (D_\mu \cB^\mn)
\nonumber\\
&&
+\si_3 e B^\kl \cB^\mn R_\klmn
+\si_4 e B_\mn \cB^\mn R,
\label{sbbd}
\eea
where 
$\si_1$,
$\si_2$,
$\si_3$,
$\si_4$
are arbitrary constants.
Note that the nonminimal curvature-coupling term 
$B^{\mu\la}\cB^\nu_{\pt{\nu}\la} R_\mn$
is proportional to the $\si_4$ term in Eq.\ \rf{sbbd}
via the identity
\beq
B_\mu^{\pt{\mu}\la} \cB_{\nu\la} 
=\frac 14 g_\mn 
(B_{\al\be} \cB^{\al\be}).
\label{identity}
\eeq
Also,
the scalar $(D_\la B_\mn) (D^\mu \cB^{\nu\la})$ 
is equivalent to terms in Eq.\ \rf{sbbd}
via the integral relation
\bea
\int d^4x e \Big[ 
(D_\la B_\mn) (D^\mu \cB^{\nu\la})
+ (D_\la B^{\la}_{\pt{\la}\nu}) (D_\mu \cB^\mn)
\nonumber\\
-\frac 14 B_\mn \cB^\mn R
+\frac 12 B^\kl \cB^\mn R_\klmn 
\Big] = 0.
\label{intrel2}
\eea

Terms quadratic in the dual tensor $\cB_\mn$
can also be considered,
including ones involving linear curvature couplings.
However,
all such terms are equivalent
to combinations of ones 
in the parity-even Lagrange density \rf{sbb}. 
Some useful identities are: 
\bea
(D_\la \cB_\mn) (D^\mu \cB^{\nu\la}) &=& 
\half (D_\la B_\mn) (D^\la B^\mn) 
\nonumber\\
&&
- (D_\la B^\la_{\pt{\la}\nu}) (D_\mu B^\mn) ,
\nonumber\\
\cB^\kl \cB^\mn R_\klmn &=& 
-B^\mn B^{\ka\la} R_{\mu\nu\ka\la}
+4B^\ln B^\mu_{\pt{\mu}\nu} R_\lm
\nonumber\\
&&
-B^\mn B_\mn R,
\nonumber\\
\cB^{\mu\la} \cB^\nu_{\pt{\nu}\la}
&=& B^{\mu\la} B^\nu_{\pt{\nu}\la}  
- \half g^\mn B^{\al\be} B_{\al\be}.
\label{idents}
\eea
Using these identities,
$\cl_{BB}$ can be rewritten as an expression 
involving terms quadratic in $\cB_\mn$.

One generalization of the above construction
involves replacing the constants $\ta_1$, $\ldots$, $\si_4$
by arbitrary functions of $B_\mn$.
This idea has recently been used 
to identify an extension to the class
of gravitationally coupled vector theories 
known as bumblebee models
\cite{mds}. 
A similar extension of the models discussed
in this work may also exist.

\subsection{Potential term}
\label{Potential term}

The term $e\cl_V$ in the action \rf{totact}
incorporates the potential $V$
triggering spontaneous symmetry breaking.
We assume that $V$ drives the formation of 
a nonzero vacuum value
\bea
\vev{B_\mn} &=& b_\mn,
\label{vev}
\eea
which breaks local Lorentz and diffeomorphism symmetry.
This implies a vacuum value for the dual field $\cB_\mn$,
\bea
\vev{\cB_\mn} &=& \cb_\mn \equiv \half\ep_\klmn b^\kl.
\eea
In general,
the potential $V$ could include dependence
on $B_\mn$, 
on covariant derivatives of $B_\mn$,
on the Levi-Civita tensor $\ep_\klmn$,
and on the metric $g_\mn$.
A pure-derivative potential has been investigated
for the vector field in certain bumblebee models
\cite{au},
and an analogous treatment could be considered here. 
However, 
for simplicity
we disregard derivative couplings in $V$
in this work.

Since the Lagrange density is an observer scalar density,
the dependence of the potential $V$ on $B_\mn$ 
can arise only through the invariants
$B_\mn B^\mn$ and $B_\mn \cB^\mn$.
Note that neither of these terms 
is invariant under the gauge transformation \rf{gaugetr}. 
Following the approach of Ref.\ \cite{bfk},
we introduce 
\bea
X_1 &\equiv& B_\mn B^\mn - x_1,
\nonumber\\
X_2 &\equiv& B_\mn \cB^\mn - x_2,
\label{x1x2}
\eea
and we write the potential as 
\beq
V=V(X_1,X_2),
\label{pot}
\eeq
where $\vev{V}=0$ is assumed.
In Eq.\ \rf{x1x2},
$x_1$ and $x_2$ are two real numbers 
representing the vacuum values of the invariants,
\bea
x_1 &\equiv& \vev{B_\mn B^\mn} 
= \vev{g^\km}\vev{g^\ln}b_\kl b_\mn,
\nonumber\\
x_2 &\equiv& \vev{B_\mn \cB^\mn} 
= \vev{g^\km}\vev{g^\ln}b_\kl \cb_\mn,
\eea
where $\vev{g^\mn}$ is the vacuum value 
of the inverse metric.

For certain purposes,
it is convenient 
to split $b_\mn$ into the independent components 
$b_{0j}$ and $b_{jk}$
and to introduce spatial vectors $\vec e$ and $\vec b$
defined by 
\bea
e^j = - b_{0j}, \quad 
b^j = \frac 12 \ep^{jkl} b_{kl},
\label{vecs}
\eea
in analogy with the separation
of the antisymmetric field strength 
into electric and magnetic vector fields
in Maxwell electrodynamics.
Under some circumstances,
it is also convenient
to perform observer rotation and boost transformations
to attain a special observer frame in which
$b_\mn$ takes a simple block-diagonal form.
This can be achieved
in a local Lorentz frame in Riemann spacetime
or everywhere in Minkowski spacetime.
Provided at least one of $x_1$ and $x_2$ is nonzero, 
the special form can be chosen as
\bea
b_\mn = \left( 
\begin{array}{cccc} 
0 &-a & 0 & 0 \\
a & 0 & 0 & 0 \\
0 & 0 & 0 & b \\
0 & 0 & -b & 0 \\
\end{array} 
\right),
\label{special}
\eea
where $a$ and $b$ are real numbers.
In this special frame,
$\vec e= (a,0,0)$ and $\vec b = (b,0,0)$,
while 
$x_1 = -2(a^2-b^2)$ and $x_2 = 4 a b$.
If both $x_1$ and $x_2$ vanish,
then the replacements 
$b_{23}= -b_{32} \to 0$, 
$b_{13}= -b_{31} \to -a$ 
can be implemented in the above block-diagonal form instead.
Note that $b_\mn$ in the special frame is determined 
by no more than two nonzero real numbers,
an improvement over the six real numbers 
required for the generic case.
However,
most of the analysis in this work makes no assumptions
about the specific form of the vacuum value $b_\mn$.

Adopting for the potential $V$ the partial derivative notation 
\bea
V_m = \fr {\prt V}{\prt X_m}, 
\quad
V_{mn} = \fr {\prt^2 V}{\prt X_m \prt X_n},
\quad
\ldots
\label{partials}
\eea
with $m,n,\ldots = 1,2,\ldots$,
the extremal conditions determining the vacuum are 
\beq
V_m = 0
\quad 
{\rm (vacuum~condition)}.
\label{vacv}
\eeq
Since $X_m=0$ in the vacuum,
the potential can be expanded 
about the vacuum as the series
\beq
V(X_1,X_2) = \half \la_{mn} X_m X_n 
+ \frac 1 6 \la_{mnp} X_m X_n X_p + \ldots ,
\label{potexp}
\eeq
where
\beq
\la_{mn} = V_{mn}(0,0),
\quad
\la_{mnp} = V_{mnp}(0,0),
\quad \ldots
\label{consts}
\eeq
are constants.
A simple example of this type
is provided by the smooth diagonal quadratic form
with only $\la_{11}$ and $\la_{22}$ nonzero.
Note that the values of the constants \rf{consts} are relevant 
to the issue of overall stability of a given vacuum,
which in general is an involved question 
\cite{kle}
and as yet remains only partially resolved 
even for comparatively simple vector-based models
\cite{stability}.
Another useful class of potentials involves
linear or quadratic Lagrange multipliers
\cite{bfk}.
However,
for most of the analysis to follow,
specifying the form of $V$ is unnecessary.

In the theory \rf{totact},
the field excitations of primary interest 
are the fluctuations in $g_\mn$ and $B_\mn$.
The metric fluctuation $h_\mn$ is given by
\beq
g_\mn = \vev{g_\mn} + h_\mn,
\label{ling}
\eeq
where $\vev{g_\mn}$ is the vacuum metric,
while $B_\mn$ can be expanded as
\beq
B_\mn = b_\mn + \Btw_\mn.
\label{linb}
\eeq
Note that the alternative expansion
$B^\mn = \vev{B^\mn} + \widetilde B^\mn$
could in principle be adopted instead
\cite{bfk}.

In Minkowski spacetime 
or in an asymptotically flat background,
we can choose coordinates with 
\beq
\vev{g_\mn} = \et_\mn 
\quad
({\rm asymptotically~flat}).
\label{asyflat}
\eeq
For these cases,
it is often convenient to introduce 
the simplifying assumption
\beq
\prt_\la b_\mn =0.
\label{vacder}
\eeq
This preserves translation invariance
and hence conservation of energy and momentum 
for the fluctuation fields $h_\mn$ and $\Btw_\mn$.
It also implies all solitonic solutions are disregarded.
Note that imposing the conditions \rf{asyflat} and \rf{vacder} 
removes most of the freedom associated with 
observer general coordinate transformations.

The excitations $h_\mn$ and $\Btw_\mn$
contain a total of 16 modes.
The explicit form of the action \rf{totact}
is required to establish their complete nature and behavior,
including whether they are NG, massive, gauge, or spectator modes,
whether they propagate or are auxiliary,
and whether the alternative Higgs mechanism occurs
\cite{bfk}.

The NG modes can be identified as the field excitations 
that preserve the minimum of the potential.
They are therefore solutions of the conditions
\beq
X_1=X_2=0
\quad
{\rm (NG~modes)}.
\label{ngcond}
\eeq
Assuming both conditions are imposed by the theory,
these represent two independent constraints to be satisfied 
by the six possible virtual Lorentz excitations of $\Btw_\mn$.
There could therefore in principle be 
as many as four Lorentz NG modes in the theory.
Determining which ones propagate 
as physical massless excitations is of definite interest
because such modes represent long-range forces 
and can therefore be expected to have 
phenomenological implications. 
Even if the spontaneous Lorentz breaking
occurs at a large scale such as the Planck mass,
resulting in suppressed massive modes at low energies,
the propagating massless modes 
can be expected to play a significant role
in the physics.
In effect,
the propagating Lorentz NG modes
form the smallest unit of the field $B_\mn$  
carrying relevant dynamical meaning at all scales.
We refer to them as `phon' modes,
a terminology adapted from phoneme, 
which is a smallest unit of language capable 
of carrying meaning.

In a given model with spontaneous Lorentz breaking 
triggered by an antisymmetric 2-tensor field,
determining the number and properties of phon modes
is key to establishing the physical content
and phenomenological implications
of the theory.
This parallels the situation for theories
with spontaneous Lorentz violation 
triggered by a vector or a symmetric 2-tensor,
where the NG modes can play a variety
of phenomenologically important roles.
For example,
certain gravitationally coupled vector theories 
with spontaneous Lorentz violation
known as bumblebee models
reproduce the Einstein-Maxwell equations in a fixed gauge,
with the NG modes identified as photon modes 
\cite{ks,bk}.
Similarly,
in a suitable theory for a symmetric 2-tensor 
generating spontaneous Lorentz violation,
the NG modes obey the nonlinear Einstein equations in a fixed gauge 
and can be identified as gravitons 
\cite{cardinal,ctw}.
Composite gravitons have been proposed 
as NG modes of spontaneous Lorentz violation
arising from self couplings of 
vectors
\cite{vector},
fermions
\cite{fermion}, 
or scalars
\cite{scalar},
following related ideas for photons
\cite{photons}.
In some models,
the NG modes can also be interpreted as
a new spin-dependent interaction
\cite{ahclt}
or as various new spin-independent forces
\cite{kt},
while in others they can generate torsion masses 
via the Lorentz-Higgs effect
\cite{bk}. 

In what follows,
we show that 
certain theories with spontaneous Lorentz breaking 
triggered by an antisymmetric 2-tensor field
contain a phon mode behaving like a scalar.
Since the phon can have nonminimal gravitational couplings,
one intriguing possibility is that it 
could play a cosmological role.
Cosmologically varying scalars can produce
Lorentz violation associated with varying couplings
\cite{klp},
and we can anticipate that phon modes 
could play the cosmological roles 
of the inflaton associated with inflation 
or the various scalar modes proposed 
to underly dark energy.
Details of these and other possible phenomenological roles 
for the phon modes is an interesting topic
for future study.

In contrast to the NG modes,
the massive modes are excitations 
increasing the value of the potential $V$ above its minimum.
It follows that there are two massive modes,
which can be identified with $X_1$ and $X_2$
or with linear combinations of these quantities.
The explicit form of $X_1$ and $X_2$ 
in terms of $h_\mn$ and $\Btw_\mn$
can be found using Eq.\ \rf{x1x2},
and their mass matrix is $\la_{mn}$.
These modes can also play a phenomenological role.
In gravitationally coupled bumblebee theories,
the massive modes modify the Newton gravitational potential
\cite{bfk},
and even modes with large masses are likely to affect
cosmological dynamics in the very early Universe. 
Analogous possibilities can be expected to arise 
for the massive modes $X_1$ and $X_2$.

\section{Minimal model}
\label{Minimal model}

This section discusses some aspects of models
with a gauge-invariant kinetic term.
The limit of Minkowski spacetime,
with the conditions \rf{asyflat} and \rf{vacder} satisfied,
is considered first.
For this purpose,
we adopt the minimal Lagrange density
\bea
\cl^{\rm min}_{B,V} &=& 
-\frac 1{12} H_\lmn H^\lmn - V,
\label{minimal}
\eea
and examine its content 
for various choices of $V$. 
We then consider some simple extensions,
including minimal current and curvature couplings.

\subsection{Minkowski spacetime}

For the analysis,
a first-order form of the Lagrange density \rf{minimal}
is useful.
Introduce a vector field $A_\mu$
with field strength and its dual given by 
\bea
F_\mn &=& \prt_\mu A_\nu - \prt_\nu A_\mu,
\nonumber\\
\cF_\mn &\equiv& \half \ep_{\mu\nu\ka\la} F^{\ka\la}.
\eea
Then,
$\cl^{\rm min}_{B,V}$
is equivalent to the first-order Lagrange density 
\beq
\cl^{\rm min}_{A,B,V} = 
\cH_\mu A^\mu - \half A_\mu A^\mu - V
\label{achdual}
\eeq
because the field $A_\mu$ is auxiliary 
and can be removed from the action,
whereupon use of the identity 
$H_\lmn H^\lmn = -6\cH_\mu \cH^\mu$
recovers $\cl^{\rm min}_{B,V}$.
Note that this procedure applies also to the path integral,
so the equivalence holds at the quantum level.
Partial integration on the first term
shows that Eq.\ \rf{achdual}
can also be written as
\beq
\cl^{\rm min}_{A,B,V} = 
\half B_\mn \cF^\mn - \half A_\mu A^\mu - V.
\label{abvdual}
\eeq
In this Lagrange density,
which is also equivalent to the minimal theory \rf{minimal},
no derivatives act on the field $B_\mn$.

Consider first the special case 
of the minimal model \rf{minimal}
with $V=0$.
The resulting Lagrange density $\cl^{\rm min}_{B,0}$
is known to describe a free scalar field
\cite{op}.
One way to see this is using
the equivalent first-order form 
\cite{svn}.
With $V=0$ in Eq.\ \rf{abvdual},
the field $B_\mn$ acts as a Lagrange multiplier
to enforce $\cF_\mn = 0$.
In Minkowski spacetime,
this implies the identity $A_\mu \equiv \prt_\mu \ph$.
Substitution yields
\beq
\cl^{\rm min}_{A,B,0} = 
-\half \prt_\mu \ph \prt^\mu \ph,
\label{freescalar}
\eeq
which is the Lagrange density for a free scalar field. 

Next,
suppose a mass term is added for the field $B_\mn$,
so that $V = m^2 B_\mn B^\mn/4$. 
The resulting Lagrange density \rf{minimal}
is known to describe a massive vector field
\cite{op}.
This can also be seen from the first-order form,
which becomes 
\beq
\cl^{\rm min}_{A,B,V} = 
\half B_\mn \cF^\mn - \half A_\mu A^\mu - \frac 1 4 m^2 B_\mn B^\mn.
\label{abmdual}
\eeq
The presence of the mass term means that $B_\mn$ 
now plays the role of an auxiliary field rather than 
a Lagrange multiplier.
Removing $B_\mn$ from the action
and using the identity $\cF_\mn \cF^\mn = - F_\mn F^\mn$
gives
\beq
m^2 \cl^{\rm min}_{A,B,V} = 
-\frac 1 4 F_\mn F^\mn - \half m^2 A_\mu A^\mu,
\label{amassive}
\eeq
which is the Lagrange density for a massive vector field. 

In the context of the present work,
we are interested in the content of the theory \rf{minimal}
when the potential $V$ takes a form that
triggers spontaneous Lorentz breaking. 
For illustrative purposes,
consider a potential $V=V(X_1)$
with nonzero quadratic coefficient $\la=\la_{11}$.
This potential depends only on $X_1$,
so the discussion in the previous section
implies at most one massive mode can be expected.

Implementing the expansion \rf{linb},
the Lagrange density becomes
\bea
\cl^{{\rm min}}_{B,V} &=& 
-\frac 1{12} \scroodle{H}_\lmn \scroodle{H}^\lmn 
- V
\nonumber\\
&=&
-\frac 1{12} \scroodle{H}_\lmn \scroodle{H}^\lmn 
- \half \la X_1^2 - \ldots, 
\label{lagssb}
\eea 
which yields the equations of motion
\beq
\prt_\la \scroodle{H^\lmn} \approx 4 \la b^\mn X_1 .
\label{eqmotmk}
\eeq 
The equivalent first-order Lagrange density is
\bea
\cl^{{\rm min}}_{B,V} &\equiv&
\half \Btw_\mn \cF^\mn - \half A_\mu A^\mu - V
\nonumber\\
&\approx&
\half \Btw_\mn \cF^\mn - \half A_\mu A^\mu 
- 2 \la (b_\mn \Btw{}^\mn)^2 .
\label{linminimal}
\eea
In the last expression,
only the leading-order term in $\Btw_\mn$
for the potential $V$ is displayed.
This is a mass term, involving the mass matrix
\beq
m_{\kl\cdot\mn} = 8 \la b_\kl b_\mn.
\eeq
However,
only the single linear combination $b_\mn \Btw{}^\mn$
of the six independent excitations in $\Btw_\mn$
is affected by this term.
For example,
in the special observer frame given by Eq.\ \rf{special},
the mass is associated with
a linear combination of $\Btw_{01}$ and $\Btw_{23}$,
so these field components
determine the massive-mode content of the theory.
The other modes remain massless.
This shows that the situation with spontaneous Lorentz breaking
is intermediate between the two Lorentz-invariant cases
with zero mass and with a conventional mass term.

The presence of the vacuum value $b_\mn$ 
defines an orientation in the theory 
that can be used for projection.
Assuming $x_1\neq 0$,
we introduce for an antisymmetric 2-tensor $T_\mn$
the orthogonal projections
\bea
T_{\| \mn} &=& \frac 1 {x_1} b_\kl T^\kl b_\mn,
\nonumber\\
T_{\perp\mn} &=& T_\mn -  T_{\| \mn} .
\label{tproj}
\eea
With this notation,
the Lagrange density \rf{linminimal}
can be written as
\bea
\cl^{{\rm min}}_{B,V} &\approx&
\half \Btw_{\perp\mn} \cF_\perp^\mn 
+\half \Btw_{\| \mn} \cF_\|{}^\mn 
\nonumber\\
&&
- \half A_\mu A^\mu 
- 2 \la x_1 \Btw_{\| \mn} \Btw_\|{}^\mn.
\label{lagproject}
\eea
This form displays explicitly the intermediate nature
of the minimal model with spontaneous Lorentz violation.
In the expression \rf{lagproject},
the projection $\Btw_{\perp\mn}$ 
is a Lagrange-multiplier field
that acts to impose the condition 
\beq
\cF_{\perp\mn} \approx 0
\label{fperp}
\eeq
in parallel with the situation when $V=0$.
However,
the projection $\Btw_{\| \mn}$ is a massive auxiliary field
obeying
\beq
\Btw_{\| \mn} \approx \fr 1 {8 \la x_1} \cF_{\|\mn},
\label{fpar}
\eeq
in analogy with the case leading to Eq.\ \rf{amassive}.
We see that the term proportional to $A_\mu A^\mu$ 
in the Lagrange density \rf{lagproject}
plays a double role,
with some combinations of the components of $A_\mu$ 
generating kinetic terms for massless NG modes
while others form a mass term for the massive mode. 

At leading order in $\Btw_\mn$,
the solutions to Eq.\ \rf{fperp} contain 
the massless NG modes in the theory,
while the massive-mode content lies in the complement
\rf{fpar}.
However,
examination of Eq.\ \rf{eqmotmk} 
reveals that no massive mode propagates
at leading order.
For example,
taking a derivative of Eq.\ \rf{eqmotmk} gives 
$b^\mn \prt_\mu X_1 \approx 0$,
and working in the special frame \rf{special}
with nonzero $x_1$ and $x_2$
shows that $X_1$ is a constant.
The result \rf{fpar} implies
$\cF_{\|\mn} \propto \Btw_{\| \mn} \propto X_1 b_\mn$,
so it follows that $\cF_{\|\mn}$ is a constant.
Adopting natural boundary conditions with $X_1=0$, 
we obtain $\cF_\mn = 0$
and hence $A_\mu = \prt_\mu \ph$.
At leading order in $\Btw_\mn$
and with these boundary conditions,
the Lagrange density \rf{lagproject}
therefore reduces to a theory of the form \rf{freescalar}
describing a single free phon mode $\ph$.
In this limit,
we see that the phon mode is the analogue 
of the scalar associated with the massless notoph
or Kalb-Ramond field
\cite{op,mkpr}. 

Further insight can be obtained
by performing a time-space decomposition on $\Btw_\mn$.
Define 
\beq
\Btw_{0j}=-\Si^j,
\qquad
\Btw_{jk} = \ep_{jkl} \Xi^l,
\label{timespacedecomp}
\eeq
in analogy with the electrodynamic decomposition 
of the field strength 
into its electric and magnetic 3-vector fields.
In terms of $\vec\Si$ and $\vec\Xi$,
the Lagrange density \rf{minimal} becomes
\bea 
\cl^{\rm min}_{B,V} &=& 
\half (\dot{\vec\Xi} + \vec\nabla\times\vec\Si)^2
-\half (\vec\nabla\cdot\vec\Xi)^2 
-V (\vec\Si, \vec\Xi).
\quad
\label{newminimal}
\eea
This form of the theory reveals that 
the only dynamical object is $\vec\Xi$,
while $\vec\Si$ is auxiliary.
It follows that at most three propagating modes
can appear in the minimal model.

For $V=0$,
use of the Helmholtz decompositions 
$\vec\Si=\vec\Si_t + \vec\Si_l$ 
and 
$\vec\Xi=\vec\Xi_t + \vec\Xi_l$ 
into divergence-free transverse and curl-free longitudinal parts
reveals the expected result
that the curl-free single degree of freedom $\vec\Xi_l$
propagates a free scalar field,
while the other fields are gauge or decouple. 
If instead the potential $V$ is a conventional mass term,
the three propagating modes are those of a massive vector. 
In contrast,
for the case of interest here
with $V$ triggering spontaneous Lorentz violation,
at most two of the six modes in $\Btw_\mn$ can be massive.
For example,
working in the special frame \rf{special},
the potential in the illustrative model \rf{lagssb} becomes
\beq
V(\vec\Si,\vec\Xi) \approx 8 \la (a\Si^1 - b \Xi^1)^2
\label{sixiv}
\eeq
at leading order in $\Btw_\mn$
and hence in $\vec\Si$ and $\vec\Xi$. 
This generates a mass matrix 
for the components $\Si^1$ and $\Xi^1$,
with the linear combination $a\Si^1 - b \Xi^1$
representing the massive mode.
The field $\Si^1$ is auxiliary.
Although $\Xi^1$ could in principle be dynamical,
it is nonpropagating at leading order in $\Btw_\mn$.
Of the remaining two degrees of freedom in $\vec\Xi$,
one is the free phon mode,
while the other can be removed 
by the residual gauge freedom
that leaves invariant the potential \rf{sixiv}.

Analogous results are obtained for the minimal model \rf{minimal}
with more general potential 
$V=V(X_1,X_2)$.
There can be up to two massive modes,
with masses determined by the eigenvalues 
of the mass matrix for $X_1$ and $X_2$.
In the special frame \rf{special},
$X_1$ and $X_2$ take the form
\bea
X_1 &=&
- 4 a \Si^1 + 4 b \Xi^1 
- 2 \vec \Si^2 + 2 \vec \Xi^2,
\nonumber\\
X_2 &=&
- 4 b \Si^1 - 4 a \Xi^1 
- 4 \vec \Si \cdot \vec \Xi.
\eea
Combinations of $\Si^1$ and $\Xi^1$ 
therefore represent the massive modes in the theory.
The field $\vec\Si$ is auxiliary
and can be eliminated from the Lagrange density,
at least in principle,
leaving only one massive degree of freedom.
As before,
this massive mode is nonpropagating at leading order.
The issue of whether it propagates at higher orders 
is an interesting open question 
but lies beyond our present scope.
This may most conveniently be addressed 
via the hamiltonian formulation
and the Dirac procedure for constraints
\cite{dirac}.

\subsection{Currents and curvature}
\label{Currents and curvature}

Next,
consider an extension of the theory \rf{minimal}
to include a coupling to a current $j_B^\mn$,
either specified externally
or formed from fields other than $B_\mn$.
The relevant Lagrange density becomes 
\bea
\cl^{\rm min}_{B,V,j} &=& 
-\frac 1{12} H_\lmn H^\lmn 
- V - \half B_\mn j_B^\mn.
\label{jminimal}
\eea
A gauge transformation of the form \rf{gaugetr}
changes $\cl^{\rm min}_{B,V,j}$ by an amount 
\bea
\de\cl^{\rm min}_{B,V,j} &=& 
\La_\nu \prt_\mu (j_V^\mn + j_B^\mn),
\qquad
\label{gtlag}
\eea
where the potential current $j_V^\mn$ is defined as 
\beq
j_V^\mn = 4V_1 B^\mn + 4V_2 \cB^\mn.
\label{potcurr}
\eeq
The result \rf{gtlag} represents the obstruction 
to gauge invariance in the theory.
Off-shell invariance is achieved whenever
the sum of the massive-mode and the matter currents
is conserved off shell.
This occurs,
for example,
if the potential $V$ vanishes
and the current $j_B^\mn$ is independently conserved.

In the present context
with spontaneous breaking of Lorentz symmetry,
the potential $V$ is nonvanishing
so gauge invariance is generically lost.
However,
the NG modes in the theory 
satisfy the conditions \rf{vacv},
so $j_V^\mn$ vanishes in this sector.
The current $j_V^\mn$ is therefore 
associated with the massive modes.
Moreover,
in parallel with the case of classical electrodynamics,
it is reasonable to take the current $j_B^\mn$ 
to be independently conserved in this sector,
\beq
\prt_\mu j_B^\mn = 0
\quad
{\rm (NG~sector)},
\label{ngcons}
\eeq 
when the massive modes are constrained to zero.
It follows from Eq.\ \rf{gtlag}
that the NG sector 
is off-shell gauge invariant
under the residual gauge transformations
satisfying the conditions \rf{ngcond}.
Also,
if $j_B^\mn$ is specified externally,
then it is conserved even in the presence of massive modes.
However,
if $j_B^\mn$ is constructed from other fields,
then it may be affected by the excitation of massive modes,
whereupon conservation may fail.

Related results emerge on shell.
The equation of motion for $\Btw_\mn$ is
\beq
\prt_\la H^\lmn = j_V^\mn + j_B^\mn.
\eeq 
This implies that the total current is conserved on shell,
\beq
\prt_\mu (j_V^\mn + j_B^\mn) = 0.
\label{currentcons}
\eeq 
It follows that the variation 
$\de\cl^{\rm min}_{B,V,j}$
of the Lagrange density vanishes on shell, 
so the gauge-symmetry breaking is an off-shell effect.

The first-order form of the theory \rf{jminimal}
can be written as the Lagrange density
\beq
\cl^{\rm min}_{A,B,V,j} = 
\half B_\mn \cF^\mn - \half A_\mu A^\mu 
- V - \half B_\mn j_B^\mn,
\label{abvjfirstorder}
\eeq
from which the original theory \rf{jminimal} 
can be recovered by partial integration on the first term
followed by elimination of the auxiliary field $A_\mu$,
as before. 
In what follows,
it is convenient to perform
a time-space decomposition for $j_B^\mn$
paralleling the decomposition \rf{timespacedecomp}.
We introduce vectors $\vec J$, $\vec K$ as
\beq
j_B^{0j}=J^j,
\qquad
j_B^{jk} = \ep^{jkl} K_l.
\label{jtimespacedecomp}
\eeq
Current conservation \rf{ngcons} implies
\beq
\vec \nabla \cdot \vec J = 0,
\quad
\vec \nabla\times\vec K - \dot{\vec J} = 0.
\eeq
Note these equations are equivalent 
to the homogeneous Maxwell equations
for the pair $(\vec E, \vec B) = (\vec K, -\vec J)$.

Consider first the Lagrange density \rf{abvjfirstorder}
with $V=0$.
Then,
$B_\mn$ remains a Lagrange-multiplier field as before,
but the associated constraint becomes 
$\cF_\mn = j_{B\mn}$.
The solution is $A_\mu = \al_\mu + \prt_\mu\ph$,
where $\al_\mu$ is the 4-vector potential 
associated with the Maxwell electromagnetic fields 
$(\vec E, \vec B) = (\vec K, -\vec J)$.
Substitution yields
\beq
\cl^{\rm min}_{A,B,V,j} = 
- \half \prt_\mu\ph \prt^\mu\ph + \ph j_\ph,
\label{abvjdual}
\eeq
where $j_\ph = \prt_\mu\al^\mu$.
A term proportional to $\al_\mu \al^\mu$
that is irrelevant for the dynamics of $\ph$ 
has been dropped.
This theory describes a scalar field $\ph$ interacting
with the current $j_\ph$.

If instead the potential in the Lagrange density \rf{abvjfirstorder}
is the mass term $V = m^2 B_\mn B^\mn/4$, 
then $B_\mn$ is auxiliary.
Removing it from the action yields 
\beq
m^2 \cl^{\rm min}_{A,B,V,j} = 
-\frac 1 4 F_\mn F^\mn - \half m^2 A_\mu A^\mu
- A_\mu j_A^\mu,
\label{massabvjfirstorder}
\eeq
where $j_A^\mu = \ep^{\al\be\ga\mu}\prt_\al j_{B\be\ga}/2$.
This equation omits the quadratic current-coupling term 
$j_B^\mn j_{B\mn}/4$,
which is irrelevant for the dynamics of $A_\mu$.
The Lagrange density \rf{massabvjfirstorder}
describes a massive vector field $A_\mu$
interacting with the current $j_A^\mu$.

For the case of interest here
with $V$ spontaneously breaking Lorentz symmetry,
we again find the theory contains a mixture  
of phon and massive modes.
However,
these modes interact with currents,
and additional SME-type couplings can appear.
Consider,
for example,
the illustrative model with $V=V(X_1)$
approximated by a quadratic term with coefficient $\la$.
Projecting the perpendicular and parallel components
of $B_\mn$, $\cF_\mn$, and $j_B^\mn$
according to Eq.\ \rf{tproj}
and substituting into Eq.\ \rf{abvjfirstorder}
gives the Lagrange density 
\bea
\cl^{\rm min}_{A,B,V,j} &\approx& 
\half \Btw_{\perp\mn} 
(\cF_\perp^\mn - j_{B\perp}^\mn) 
\nonumber\\
&&
+\half \Btw_{\|\mn} 
(\cF_\|^\mn - j_{B\|}^\mn) 
- 2 \la x_1 \Btw_{\| \mn} \Btw_\|{}^\mn
\nonumber\\
&&
- \half A_\mu A^\mu 
- \half b_\mn j_B^\mn.
\label{jlagproject}
\eea
This reveals that the projection $\Btw_{\perp\mn}$ 
is a Lagrange-multiplier field imposing the constraint 
\beq
\cF_{\perp\mn} \approx j_{B\perp}^\mn ,
\label{jfperp}
\eeq
while the projection $\Btw_{\|\mn}$ is an auxiliary field
given by
\beq
\Btw_{\| \mn} \approx \fr 1 {8 \la x_1} 
(\cF_\|^\mn - j_{B\|}^\mn) .
\label{jfpar}
\eeq
As before,
the NG modes are contained 
in the solutions to Eq.\ \rf{jfperp},
while the massive-mode content is in the complement
\rf{jfpar}
and is constrained by current conservation.
Adopting natural boundary conditions 
for the equations of motion again leads to $X_1=0$.
The solution for $A_\mu$ 
can be written as $A_\mu = \al_\mu + \prt_\mu\ph$,
where $\ph$ is the phon mode and
$\al_\mu$ is the 4-vector potential 
for the Maxwell fields 
$(\vec E, \vec B) = (\vec K, -\vec J)$.
At leading order,
the only propagating mode is the phon.
Removing the Lagrange-multiplier and auxiliary modes from the theory
yields the Lagrange density 
\beq
\cl^{\rm min}_{A,B,V,j} \approx 
- \half \prt_\mu\ph \prt^\mu\ph + \ph j_\ph
- \half b_\mn j_B^\mn
- \half \al_\mu \al^\mu,
\label{phonabvjdual}
\eeq
where $j_\ph = \prt_\mu \al^\mu$. 
This describes an interacting phon 
along with an SME-type coupling to the current $j_B^\mn$
and an induced current-current coupling.

Another interesting extension of the minimal theory \rf{minimal}
is obtained in passing from Minkowski to Riemann spacetime
and adding the Einstein-Hilbert term \rf{eh}.
The relevant Lagrange density is
\bea
e\cl^{\rm min}_{R,B,V,j} &=& 
\fr e{2\ka} (R - 2\La)
- \frac 1{12} e H_\lmn H^\lmn 
\nonumber\\
&&
- eV - \half e B_\mn j_B^\mn.  
\label{Rminimal}
\eea
Much of the discussion in Minkowski spacetime remains valid,
but some derivations face obstructions.

The introduction of the vector field $A_\mu$
and the construction of the equivalent first-order form
\bea
e\cl^{\rm min}_{R,A,B,V,j} &=& 
\fr e{2\ka} (R - 2\La)
+ \half e B_\mn \cF^\mn 
\nonumber\\
&&
- \half e A_\mu A^\mu 
- eV - \half e B_\mn j_B^\mn 
\label{Rabvjfirstorder}
\eea
proceeds as before because the derivatives in $H_\lmn$ 
can be taken as covariant and hence 
the partial integration performed. 
Global statements obtained from the Lagrange density 
become local statements,
including the equations of motion
and the results for the current $j_B^\mn$.

If $V=0$
and the topology of the spacetime manifold $M$ is trivial,
the theory describes a scalar field in Riemann spacetime.
However,
the solution for $A_\mu$ that leads to this interpretation
is valid only locally 
if the first cohomology class $H^1(M,\mathbb{R})$ 
is nonvanishing.
This issue is absent 
if $V$ is taken as the mass term $V = m^2 B_\mn B^\mn/4$, 
when the theory describes a massive vector
in Riemann spacetime.

If instead $V$ triggers spontaneous Lorentz breaking,
the vacuum value $b_\mn$ can vary with spacetime position
and hence have nontrivial derivative
in the general case
\cite{akgrav}.
The field strength $H_\lmn$ can therefore
acquire a nonzero contribution
even when $\Btw_\mn$ vanishes. 
However,
this has no effect on the first-order form \rf{Rabvjfirstorder}.
For example,
performing the decomposition \rf{tproj}
for the illustrative model with $V=V(X_1)$
approximated by a quadratic term
yields the Lagrange density 
\bea
e\cl^{\rm min}_{R,A,B,V,j} &\approx& 
\fr e{2\ka} (R - 2\La)
+ \half e\Btw_{\perp\mn} 
(\cF_\perp^\mn - j_{B\perp}^\mn) 
\nonumber\\
&&
+\half e\Btw_{\|\mn} 
(\cF_\|^\mn - j_{B\|}^\mn) 
- 2 \la x_1 e\Btw_{\| \mn} \Btw_\|{}^\mn
\nonumber\\
&&
- \half eA_\mu A^\mu 
- \half e b_\mn j_B^\mn,
\label{Rjlagproject}
\eea
where $b_\mn$ may now vary with position.
The constraint \rf{jfperp} still holds,
but if $H^1(M,\mathbb{R})$ is nontrivial
then the general solution for $A_\mu$ includes 
all independent closed nonexact 1-forms 
with suitable support on the spacetime.
The latter can be viewed as additional topological modes 
in the theory,
but since these modes are nonexact 
they cannot play the role of topological phon modes.
There is still only one phon,
which propagates in Riemann spacetime
and interacts with a current that includes 
a contribution from the topological modes.  
In the special case of an asymptotically flat spacetime
with $\La = 0$ and trivial topology,
the condition \rf{vacder} holds
and the topological modes are absent.
At leading order in $h_\mn$,
the phon then propagates in a Minkowski background
with weak-field coupling to the metric.

\section{Nonminimal model}
\label{Nonminimal model}

The effects of Lorentz violation on gravity 
can be characterized in a general way 
by constructing the effective field theory
for the metric and curvature
while allowing arbitrary Lorentz-violating couplings
\cite{akgrav}.
This procedure generates the gravity sector of the SME
in Riemann spacetime.
At leading order in the curvature,
three basic types of Lorentz-violating couplings arise.
Each involves a coefficient field for Lorentz violation
that upon acquiring a vacuum value
generates a Lorentz-violating coupling for gravity.
The theory \rf{totact} for the antisymmetric 2-tensor $B_\mn$
has the interesting feature 
of containing all three kinds of couplings,
despite being comparatively simple.

In this section,
we consider a particular restriction of the theory \rf{totact} 
that suffices to exhibit all three kinds of couplings.
The theory includes some of the nonminimal curvature couplings
obtained in 
Sec.\ \ref{Kinetic term}.
Following the specification of the Lagrange density,
the equations of motion 
and energy-momentum conservation law 
are obtained.
The results are linearized
and some implications for the mode content are obtained.
We then apply the formalism of Ref.\ \cite{qbak}
to extract the post-newtonian metric. 

\subsection{Action}

At leading order in the curvature,
the three basic types of Lorentz-violating couplings
include ones to 
the traceless Ricci tensor,
the Weyl tensor,
and the scalar curvature.
The corresponding SME coefficient fields
are conventionally denoted as
$s^\mn$, $t^\klmn$, and $u$. 
We adopt here an extension of the minimal model 
of the previous section that suffices to include all three.
It is constructed by adding nonzero couplings 
of the $\ta_3$, $\ta_4$, and $\ta_5$ types
displayed in 
Sec.\ \ref{Kinetic term}.
For simplicity,
we assume $\La = 0$ and $j_B^\mn = 0$
but include a matter Lagrange density $\cL_M$
to act as a gravitational source.
The potential $V(X_1,X_2)$ 
triggering spontaneous Lorentz violation
is taken to satisfy the vacuum condition \rf{vacv}
and to have the expansion \rf{potexp}
involving the constants $\la_{mn}$.

The chosen Lagrange density can be written in the form 
\bea
e \cl^{\rm nonmin} &=& 
\fr {e}{2\ka} R 
-\frac 1{12} e H_\lmn H^\lmn 
-eV + e\cl_M
\nonumber\\
&&
+ \fr {e}{2\ka} \big(
\xi_1 B^\kl B^\mn R_\klmn
+\xi_2 B^{\ln} B^\mu_{\pt{\la}\nu} R_{\lm}
\nonumber\\
&&
\hskip30pt
+\xi_3 B^\mn B_\mn R \big).
\label{act2}
\eea
For convenience in the analysis to follow,
we have extracted a factor of $2\ka$ 
from the coupling constants 
$\ta_3$, $\ta_4$, $\ta_5$
and relabeled them as 
$\xi_1$, $\xi_2$, $\xi_3$.

At the level of the action,
the theory \rf{act2}
implies an explicit correspondence between $B_\mn$ 
and the three SME coefficient fields
$s^\mn$, $t^\klmn$, and $u$. 
We find
\bea
(s_B)^\mn &=&
(2\xi_1 + \xi_2) ( B^\mu_{\pt{\mu}\al} B^{\nu\al} 
-\frac 14 g^\mn B^{\al\be} B_{\al\be} ),
\nonumber\\
(t_B)^{\ka\la\mu\nu} &=& 
\frac 2 3 \xi_1 (
B^\kl B^\mn + \half B^\km B^\ln  - \half B^\kn B^\lm)  
\nonumber\\
&&
- \half \xi_1 ( g^{\ka\mu} B^\la_{\pt{\la}\al} B^{\nu\al}
-g^{\la\mu} B^\ka_{\pt{\ka}\al} B^{\nu\al}
\nonumber\\
&&
\hskip 20pt
-g^{\ka\nu} B^\la_{\pt{\la}\al} B^{\mu\al}
+g^{\la\nu} B^\ka_{\pt{\la}\al} B^{\mu\al} )
\nonumber\\
&&
+\frac 1 6 \xi_1 (g^{\ka\mu} g^{\la\nu}
- g^{\la\mu} g^{\ka\nu}) B^{\al\be} B_{\al\be},
\nonumber\\
u_B &=&-( \frac 1 6  \xi_1 + \frac 1 4 \xi_2 + \xi_3 )
B^{\al\be} B_{\al\be}.
\label{stua}          
\eea
The reader is cautioned that the vacuum values
of the coefficient fields implied by these equations
differ by scalings
from those that appear
in the final linearized effective Einstein equations 
\cite{qbak}.
This issue is revisited in Sec.\ \ref{pn} below.
 
The gravitational field equations 
follow from the Lagrange density \rf{act2} 
by varying with respect to $g_\mn$,
while holding $B_\mn$ and any matter fields fixed.  
Explicitly,
we find
\bea
G^\mn &=& \ka (T_M)^\mn + \ka (T_B)^\mn 
\nonumber\\
&&
+ (T_{\xi_1})^\mn + (T_{\xi_2})^\mn + (T_{\xi_3})^\mn.
\label{metricfe}
\eea
The first term on the right-hand side is the matter 
energy-momentum tensor.
The second term is 
the contribution to the energy-momentum tensor 
arising from the kinetic and potential terms for $B_\mn$.
It is given by
\bea
(T_B)^\mn &=& 
\half H^{\al\be\mu} H^{\nu}_{\pt{\nu)}\al\be}
-\frac 1{12} g^\mn H^{\al\be\ga} H_{\al\be\ga} 
- g^\mn V
\nonumber\\
&&
+ 4 B^{\al\mu} B_\al^{\pt{\al}\nu} V_1
+g^\mn {\cB_{\al\be}} B^{\al\be} V_2.
\label{tb}
\eea
The remaining three terms in Eq.\ \rf{metricfe} are due to the 
nonminimal gravitational couplings.
For the $\xi_1$ coupling,
we find
\bea
(T_{\xi_1})^\mn &=& 
\xi_1 (\half g^\mn B^{\al\be} B^{\ga\de} R_{\al\be\ga\de} 
+\frac 32 B^{\be\ga} B^{\al\mu} R^{\nu}_{\pt{\nu}\al\be\ga}
\nonumber\\
&&
+\frac 32 B^{\be\ga} B^{\al\mu} R^{\nu}_{\pt{\nu}\al\be\ga}
+ D_\al D_\be B^{\al\mu} B^{\nu\be}
\nonumber\\
&&
+ D_\al D_\be B^{\al\nu} B^{\mu\be}).
\label{t1}
\eea
The contribution from the $\xi_2$ coupling is 
\bea
(T_{\xi_2})^\mn &=& 
\xi_2 (\half g^\mn B^{\al\ga} B^\be_{\pt{\be}\ga} R_{\al\be}
- B^{\al\mu} B^{\be\nu} R_{\al\be} 
\nonumber\\
&&
-B^{\al\be} B^\mu_{\pt{\mu}\be} R^\nu_{\pt{\nu}\al}
-B^{\al\be} B^\nu_{\pt{\nu}\be} R^\mu_{\pt{\mu}\al}
\nonumber\\
&&
+\half D_\al D^{\mu} B^{\nu}_{\pt{\nu}\be} B^{\al\be}
+\half D_\al D^{\nu} B^{\mu}_{\pt{\mu}\be} B^{\al\be}
\nonumber\\
&&
-\half D^2 B^{\al\mu} B_\al^{\pt{\al}\nu}
-\half g^\mn D_\al D_\be B^{\al\ga} B^{\be}_{\pt{\be}\ga}).
\qquad
\label{t2}
\eea
Finally,
for the $\xi_3$ coupling we obtain
\bea
(T_{\xi_3})^\mn &=& 
\xi_3 (D^{\mu} D^{\nu} B^{\al\be} B_{\al\be}
-g^\mn D^2 B^{\al\be} B_{\al\be}
\nonumber\\
&&
-B^{\al\be} B_{\al\be} G^\mn
+2 B^{\al\mu} B^{\nu}_{\pt{\nu}\al} R). 
\label{t3}
\eea

The equations of motion for the antisymmetric 2-tensor 
are obtained by varying the Lagrange density \rf{act2} 
with respect to $B_\mn$,
while holding the metric and any matter fields fixed.
They can be written in the form
\beq
D_\al H^{\al\mu\nu} = j_V^\mn + j_R^\mn,
\label{bfe}
\eeq
where the potential current $j_V^\mn$ 
is given by Eq.\ \rf{potcurr}
and the curvature current $j_R^\mn$ is defined as
\bea
j_R^\mn &=&  
-\fr {2\xi_1} {\ka} B_{\al\be} R^{\al\be\mu\nu}
+\fr {\xi_2}{\ka} 
B_\al^{\pt{\al}[\mu} R^{\nu]\al}
-\fr {2 \xi_3}{\ka} B^\mn R.
\nonumber\\
\label{rbfe}
\eea
The sum of the currents is covariantly conserved on shell,
\beq
D_\mu (j_V^\mn + j_R^\mn) = 0,
\label{covcurrentcons}
\eeq
as a consequence of the minimal kinetic term for $B_\mn$ 
chosen for the theory \rf{act2}.
This result is the nonminimal analogue of Eq.\ \rf{currentcons}.
Since $j_V^\mn$ involves the derivatives $V_1$ and $V_2$,
which are nonzero when the massive modes are excited,
Eq.\ \rf{covcurrentcons} can serve as a constraint
on the massive modes.
However,
when the massive modes vanish,
it can be viewed instead as a constraint 
on the curvature.
This issue is revisited 
as part of the discussion of the linearized limit 
in Sec.\ \ref{linearization} below.

For the matter described by the Lagrange density $\cl_M$,
the equations of motion follow
by variation with respect to the matter fields.
The matter energy-momentum tensor $(T_M)^\mn$ 
is covariantly conserved,
\beq
D_\mu (T_M)^\mn = 0.
\label{cons2}
\eeq
This can be verified explicitly 
as follows.
First,
note that the components 
$(T_B)^\mn$, $(T_{\xi_1})^\mn$, 
$(T_{\xi_2})^\mn$, and $(T_{\xi_3})^\mn$
of the total energy-momentum tensor
satisfy the relation
\beq
\ka D_\mu (T_B)^\mn = 
- D_\mu [(T_{\xi_1})^\mn + (T_{\xi_2})^\mn + (T_{\xi_3})^\mn].
\label{cons1}
\eeq
This can be checked by evaluating the left-hand side 
using the field equations \rf{bfe}, 
the identity \rf{cyclic}, 
the Bianchi identities for the curvature tensor,
and the identity \rf{identity}.
Next,
take the covariant divergence 
of the gravitational field equations \rf{metricfe} 
and impose the traced Bianchi identities 
$D_\mu G^\mn =0$.
Substitution of Eq.\ \rf{cons1} then yields
the matter energy-momentum conservation law \rf{cons2}.

\subsection{Linearization}
\label{linearization}

This subsection explores
the linearized version of the theory \rf{act2}
in an asymptotically flat spacetime. 
We choose coordinates as in Eq.\ \rf{asyflat}
and impose the condition \rf{vacder}.
The weak-field limit is taken, 
so only the leading-order terms 
in the fluctuations $h_\mn$ and $\Btw_\mn$
are kept.
The fluctuations are assumed to vanish 
in the asymptotic region,
far from any matter sources.
As usual,
raising and lowering of indices 
on linear quantities 
is understood to involve the Minkowski metric.

In the minimum of the potential,
$X_1=X_2=0$ and the vacuum solution satisfies
\bea
\et^\km \et^\lm b_\kl b_\mn &=& x_1,
\quad 
\et^\km \et^\lm b_\kl \cb_\mn = x_2.
\label{vac1}
\eea
At linear order,
$X_1$ and $X_2$ take the form 
\bea
\Xxtw &\approx& 
2 b_\mn \Btw{}^\mn 
- 2 b_{\mu\al} b_{\nu}^{\pt{\nu}\al} h^\mn,
\nonumber\\
\Xytw &\approx& 
2 \cb_\mn \Btw{}^\mn 
- \half x_2 h^{\al}_{\pt{al}\al}.
\label{Xtws}
\eea
These combinations represent the massive modes in the theory
at this order.

At leading order,
the field equations for the metric 
retain the form \rf{metricfe},
but all quantities are understood to be linearized.
The linearization of the Einstein tensor 
on the left-hand side is standard.
The first term on the right-hand side is the linearized 
energy-momentum tensor for ordinary matter.
Explicit expressions for the remaining terms
on the right-hand side are
\bea
(T_B)_\mn & \approx & 4 (\Vxxb \Xxtw + \Vxyb \Xytw) 
b_{\mu\al} b_\nu^{\pt{\nu}\al} 
\nonumber\\
&&
+\et_\mn (\Vyyb \Xytw + \Vxyb \Xxtw) x_2,
\nonumber\\
(T_{\xi_1})_\mn & \approx & 
\xi_1 \big[\half \et_\mn b^{\al\be} b^{\ga\de} R_{\al\be\ga\de} 
+4b^{\be\ga} b^\al_{\pt{\al}(\mu} R_{\nu)\al\be\ga}
\nonumber\\
&&
\pt{\xi_1}
+ 2 b^\al_{\pt{\al}\mu} b^\be_{\pt{\be}\nu} R_{\al\be}
+ 4 b^\al_{\pt{\al}(\mu} \prt^\be D_\al B_{\nu)\be} \big],
\nonumber\\
(T_{\xi_2})_\mn &\approx& 
\xi_2 \big[
\et_\mn ( b^{\al\ga} b^\be_{\pt{\be}\ga} R_{\al\be} 
- \half b^{\al\ga} \prt^\be H_{\al\be\ga}
\nonumber\\
&&
- \frac 14 b^{\al\be} b^{\ga\de} R_{\al\be\ga\de}
- \half b^{\al\be} \prt^\ga D_\ga B_{\al\be} )
\nonumber\\
&&
-b^\al_{\pt{\al}\mu} b^\be_{\pt{\be}\nu} R_{\al\be}
-2 b^{\al\be} R_{\al(\mu} b_{\nu)\be}
\nonumber\\
&&
+b^{\al\ga} b^\be_{\pt{\be}\ga} R_{\al\mu\nu\be} 
-\half b^\al_{\pt{\al}(\mu} R_{\nu)\al\be\ga} b^{\be\ga}
\nonumber\\
&&
+b^{\be\ga} \prt_{(\mu} D_\be B_{\nu)\ga}
+b^\al_{\pt{\al}(\mu} \prt^\be D_{\nu)} B_{\al\be}
\nonumber\\
&&
+b^\al_{\pt{\al}(\mu} \prt^\be D_\be B_{\nu)\al} \big],
\nonumber\\
(T_{\xi_3})_\mn &\approx& 
\xi_3 [(\prt_\mu \prt_\nu -\et_\mn \prt^\al\prt_\al) \Xxtw
\nonumber\\
&&
-2 b_\mu^{\pt{\mu}\al} b_{\nu\al}R
- x_1 G_\mn].
\label{t3lin}
\eea
In these expressions, 
all covariant derivatives and curvatures 
are taken to linear order in $h_\mn$ and $\Btw_\mn$. 

The linearized field equations for the fluctuations
$\Btw_\mn$ take the form 
\beq
\prt_\al H^{\al\mu\nu} = j_V^\mn + j_R^\mn,
\label{linearb}
\eeq
where $H_\lmn$ is constructed using $\Btw_\mn$.
The linearized currents $j_V^\mn$ and $j_R^\mn$ 
are given by
\bea
j_V^\mn &=&
4 (\Vxxb \Xxtw + \Vxyb \Xytw) b^\mn 
+4 (\Vyyb \Xytw + \Vxyb \Xxtw) \cb^\mn
\nonumber\\
\eea
and
\bea
j_R^\mn &=& 
-\fr {2\xi_1}{\ka} b_{\al\be} R^{\al\be\mu\nu}
+\fr {2\xi_2}{\ka} b_\al^{\pt{\al}[\mu} R^{\nu]\al}
-\fr {2\xi_3}{\ka} b^\mn R.
\nonumber\\
\eea

The identity $\prt^\nu \prt^\mu H_{\mu\nu\la}=0$
implies that the total current is conserved,
\beq
\prt_\mu (j_V^\mn + j_R^\mn) = 0,
\label{prtcurrentcons}
\eeq
This can be interpreted as a constraint
on massive-mode excitations,
which appear in $j_V^\mn$,
and it also implies conditions on the linearized curvatures.

To investigate further,
it is convenient to introduce
two combinations of the linearized massive modes
$X_1$, $X_2$ and the linearized scalar curvature $R$
given by
\bea
\Xop &=& 4(\Vxxb \Xxtw + \Vxyb \Xytw) 
- \fr 1 {2\ka} (\xi_2+4\xi_3)R,
\nonumber\\
\Xtp &=& 4(\Vyyb \Xytw + \Vxyb \Xxtw).
\label{psis}
\eea
In terms of these variables,
the conservation law \rf{prtcurrentcons}
takes the simple form
\beq
b_{\al\nu}\prt^\al \Xop 
+\cb_{\al\nu}\prt^\al \Xtp
-\fr 1\ka (4 \xi_1+\xi_2) 
b^{\al\be} \prt_{\al} R_{\be\nu} = 0.
\label{constr5}
\eeq
By applying the differential operator
$b^\nu_{\pt{\nu}\ga} b^\ga_{\pt{\ga}\de}\prt^\de$,
which cancels the first two terms containing the massive modes,
we obtain a condition on derivatives 
of the linearized Ricci tensor,
\bea
0 &=& 
(4\xi_1 + \xi_2)
b^\nu_{\pt{\nu}\ga} b^\ga_{\pt{\ga}\de} b^{\al\be} 
\prt^\de \prt_{\al} R_{\be\nu}
\nonumber\\
&=& (4\xi_1 + \xi_2) 
b^\nu_{\pt{\nu}\ga} b^\ga_{\pt{\ga}\de} b^{\al\be} 
\prt^\de \prt_{\al} 
[ (T_M)_{\be\nu}-\half\et_{\be\nu} 
(T_M)^\mu_{\pt{\mu}\mu} ]
\nonumber\\
&&
+ (4\xi_1 + \xi_2) \times O(\xi).
\label{constr2}
\eea
In the second equation above,
we have substituted for the linearized Ricci tensor
in terms of the linearized matter energy-momentum tensor
using the gravitational field equations.

If we solve the field equations 
perturbatively in the couplings $\xi_1$, $\xi_2$,
then at lowest order  Eq.\ \rf{constr2} 
generates a direct constraint on ordinary matter.
Consider,
for example,
a static distribution of mass given by
\beq
(T_M)_\mn = \rh \de^0_{\pt{0}\mu} \de^0_{\pt{0}\nu},
\eeq
and adopt an observer coordinate system 
with $b_\mn$ lying in a configuration 
with the vectors \rf{vecs} given by
$\vec e = (a,0,0)$ and $\vec b=(b \cos \th, b \sin \th,0)$.
Then the constraint \rf{constr2} becomes
\beq
(4\xi_1+\xi_2)a^2 b\sin \th \fr {\prt^2 \rh}{\prt z \prt x}
=(4\xi_1 + \xi_2) \times O(\xi),
\label{constr3}
\eeq
which is generically inconsistent with small corrections 
to the general-relativistic behavior of matter.
In this work,
we are interested in post-newtonian corrections
to general relativity
rather than in more radical proposals.
To retain conventional properties of matter,
we therefore limit attention in what follows
to models satisfying the condition 
\beq
4 \xi_1 + \xi_2 = 0.
\label{cond}
\eeq
In these models,
the constraint \rf{constr2} is satisfied automatically.
Note that conditions of this type also arise 
in the post-newtonian limit of other theories
with nonminimal gravitational couplings,
such as vector-tensor models without a potential term 
\cite{cmw}.

Imposing the condition \rf{cond} 
eliminates the last term in the conservation law \rf{constr5},
which reduces to a constraint 
on the massive-mode combinations $\Xop$, $\Xtp$.
Assuming at least one of $x_1$ and $x_2$ is nonzero,
we can choose the special observer reference frame \rf{special} 
in which $\vec b = c \vec e$
for some nonzero real number $c$.
The conservation law \rf{constr5} then implies  
\bea
(1+c^2) \vec e \times \vec \nabla \Xop &=& 
\vec e \times \vec \nabla \ch,
\nonumber\\
\Xtp &=& \fr 1 c (\Xop - \ch),
\label{case1} 
\eea
where 
$\ch$ is a purely static function 
obeying $\vec e \cdot \vec \nabla \ch = 0$
in this special frame.
Imposing the boundary conditions 
$\ch=0$ at $t=t_0$
and $\Xop=0$ at spatial infinity
then implies 
\beq
\Xop=\Xtp=0
\label{zeromassivemodes}
\eeq
everywhere in spacetime.
This shows that the only propagating modes
in the theory \rf{act2}
subject to the consistency requirement \rf{cond} 
and to a plausible choice of boundary conditions
are gravitational and phon modes.

Other boundary conditions can also be adopted, 
for which $\Xop$ and $\Xtp$ could potentially
act as extra sources for non-massive modes 
in $h_\mn$ and $\Btw_\mn$.
A similar situation arises for the massive mode
in bumblebee models,
which under suitable boundary conditions
yields a modified Einstein-Maxwell theory 
even in the weak static limit
\cite{bfk}.
An investigation along related lines 
for the Lagrange density \rf{act2}
or the general action \rf{totact} is of interest
but lies beyond our present scope.

\subsection{Post-newtonian metric}
\label{pn}

In this subsection,
we manipulate the equations of motion
to extract a version of the linearized gravitational field equations
that depends on the vacuum values $b_\mn$
but is independent of $\Btw_\mn$.
This is achieved at leading order
in the nonminimal couplings.
A match is then made 
to the general form of the linearized gravitational field equations
obtained in Ref.\ \cite{qbak},
and the post-newtonian metric extracted.

Consider first the linearized dynamics of the fluctuations $\Btw_\mn$.
We adopt the requirement \rf{cond} 
for compatibility with conventional properties of matter
and choose boundary conditions 
yielding the condition \rf{zeromassivemodes}
on the massive modes.
The field equations \rf{linearb} then simplify
to the form
\bea
\prt^\al H_{\al\mu\nu} &=& 
\fr {\xi_2}{2\ka} (b^{\al\be} R_{\al\be\mu\nu}
+4 b^\al_{\pt{\al}[\mu} R_{\nu]\al}
+b_\mn R).
\qquad
\label{linearb2}
\eea
This result can be interpreted as an equation 
for the fluctations $\Btw_\mn$ subject to the constraints
\bea
b_\mn \Btw{}^\mn &=& b_{\mu\la} b_\nu^{\pt{\nu}\la} h^\mn
+a_1 R,
\nonumber\\
\cb_\mn \Btw{}^\mn &=& \half x_2  h^\al_{\pt{\al}\al}
+a_2 R,
\label{constr6}
\eea
where $a_1$ and $a_2$ are 
given by
\bea
a_1 &=& \left( \fr {\Vyyb}{\Vyyb \Vxxb-\Vxyb^2}\right)
\left( \fr {- \xi_1 + \xi_3}{4\ka}\right),
\nonumber\\
a_2 &=& \left( \fr {\Vxyb}{\Vxyb^2-\Vyyb \Vxxb}\right)
\left( \fr {-\xi_1 + \xi_3}{2\ka}\right).
\label{a1a2}
\eea
In these expressions,
the coupling constant $\xi_2$ has been
eliminated in favor of $\xi_1$ using
the condition \rf{cond}.

The desired goal is to use the field equations \rf{linearb2}
to eliminate all appearances of $\Btw_\mn$
in the linearized gravitational field equations \rf{metricfe},
which corresponds to eliminating $\Btw_\mn$
from the partial energy-momentum tensors \rf{t3lin}.
We work here at leading order
in the coupling constants $\xi_1$ and $\xi_3$.
A useful first step 
is to choose boundary conditions on the dynamics 
ensuring that the projection $b^\mn H_\lmn$ 
is first order in $\xi_1$.
To achieve this,
consider the cyclic identity 
\bea
&&
\hskip -10 pt
\prt^\al\prt_\al (b^\mn H_\lmn)
\nonumber\\
&&
\hskip 10 pt
= b^\mn (\prt_\mu \prt^\al H_{\al\nu\la}
+\prt_\la \prt^\al H_{\al\mu\nu}
+\prt_\nu \prt^\al H_{\al\la\mu}).
\qquad
\label{cyclic2}
\eea
Inserting the field equations \rf{linearb2}
yields
\bea
\prt^\al \prt_\al (b^\mn H_\lmn) &=& 
\fr{8\xi_1}{\ka}
( b^\al_{\pt{\al}\la}b^{\be\ga}\prt_\be G_{\ga\al}
\nonumber\\
&&
- b^{\al\be}b^\ga_{\pt{\ga}\be} \prt_\al G_{\ga\la}
+ b^{\al\be}b^\ga_{\pt{\ga}\be} \prt_\la G_{\al\ga}
\nonumber\\
&&
- \half b^{\al\be}b_{\la\be} \prt_\al R
+ \frac 1 4 x_1 \prt_\la R).
\label{linearb3}
\eea
This is a hyperbolic equation for the projection $b^\mn H_\lmn$ 
with source term of order $O(\xi_1/\ka)$,
where $\xi_1/\ka$ is taken as a small dimensionless parameter 
controlling the size of the nonminimal couplings.
We can ensure that the solutions are also of order $O(\xi_1/\ka)$,
\beq
b^\mn H_{\mu\nu\la} \sim O(\xi_1/\ka),
\label{bc2}
\eeq
by choosing boundary conditions
to eliminate the homogeneous solutions to Eq.\ \rf{linearb3}.
This choice implies that 
the projected covariant derivative $b^\mn D_\mu B_{\nu\la}$ 
is of order $O(\xi/\ka)$,
\beq
b^\mn D_\mu B_{\nu\la} \approx 
\half b^\mn H_\lmn - \half \prt_\la (a_1 R) 
\sim O(\xi/\ka).
\label{cond2}
\eeq

With these results in hand,
we can tackle the elimination of $\Btw_\mn$ 
from the partial energy-momentum tensors \rf{t3lin}.
Inspection reveals that 
the terms in the latter involving the fluctuations $\Btw_\mn$ 
either are higher order in the nonminimal couplings $\xi_1$, $\xi_3$ 
or are expressible in terms of the metric fluctuations $h_\mn$.
Some manipulation then yields 
effective linearized field equations 
for the metric fluctuations $h_\mn$
at leading order in $\xi_1$ and $\xi_3$.
In terms of linearized curvature tensors,
these equations can be expressed as 
\bea
R_\mn & \approx & 
\ka (S_M)_\mn - 2 \xi_1 b_\mu^{\pt{\mu}\al} b_{\nu\al} R
+ 6 \xi_1 b^{\al\be} b^\ga_{\pt{\ga}(\mu} R_{\nu)\ga\al\be}
\nonumber\\
&&
+ 6 \xi_1 b_\mu^{\pt{\mu}\al} b_{\nu}^{\pt{\nu}\be} R_{\al\be}
+ 8 \xi_1 b^\be_{\pt{\be}(\mu} R_{\nu)\al} b_\be^{\pt{\be}\al} 
\nonumber\\
&&
- 5 \et_\mn \xi_1 b^{\al\ga} b^\be_{\pt{\be}\ga} R_{\al\be}
+ \frac 32 \et_\mn \xi_1 b^{\al\be} b^{\ga\de} R_{\al\be\ga\de}
\nonumber\\
&&
+ 4 \xi_1 b^\al_{\pt{\al}\ga} b^{\be\ga} R_{\mu\al\nu\be}
- \xi_3 b^{\al\be} b_{\al\be} R_\mn
\nonumber\\
&&
+ \et_\mn \xi_1 b^{\al\be} b_{\al\be} R,
\label{heqns}
\eea
where $(S_M)_\mn$ is the trace-reversed energy-momentum tensor
for the matter.

At this stage,
the expression \rf{heqns} 
for the linearized gravitational field equations 
can be matched to the general form 
\beq
R_\mn = 
\ka S_\mn 
+ (\Ph^{\sb})_\mn + (\Ph^{\tb})_\mn + (\Ph^{\ub})_\mn,
\label{genform}
\eeq
obtained in Ref.\ \cite{qbak},
where the quantities on the right-hand side are defined as
\bea
\Ph^{\sb}_\mn &=& 
\half \nsy (\sb_B)^{\al\be} R_{\al\be}
-2 (\sb_B)^\al{\hskip-6pt}_{\pt{\al}(\mu} R_{\nu)\al} 
+ \half (\sb_B)_\mn R
\nonumber\\
&&
+ (\sb_B)^{\al\be} R_{\al\mu\nu\be},
\nonumber
\\
\Ph^{\tb}_\mn &=& 
2(\tb_B)^{\al\be\ga}{\hskip-15pt}_{\pt{\al\be\ga}(\mu} 
R_{\nu)\ga\al\be}
+ 2(\tb_B)^{\pt{\mu}\al\pt{\nu}\be}_{\mu\pt{\al}\nu} R_{\al\be}
\nonumber\\
&&
+ \frac 12 \nsy (\tb_B)^{\al\be\ga\de} R_{\al\be\ga\de}
\nonumber\\
&=& 0,
\nonumber\\
\Ph^{\ub}_\mn &=& \ub_B R_\mn.
\label{tphi}
\eea
Note that the net contribution to $(\Ph^{\tb})_\mn$ vanishes,
as a consequence of an identity satisfied
by the coefficients $(\tb_B)^\klmn$
\cite{qbak}.
In the expressions \rf{tphi},
the coefficients for Lorentz violation
$(\sb_B)^\mn$, $(\tb_B)^\klmn$, $\ub_B$
can be expressed explicitly in terms of the vacuum value $b_\mn$ as
\bea
(\sb_B)^\mn &=& 
2 \xi_1 (b^\mu_{\pt{\mu}\al} b^{\nu\al} - \frac 14 
\et^\mn b^{\al\be} b_{\al\be}),
\nonumber\\
(\tb_B)^{\ka\la\mu\nu} &=& 
2 \xi_1 (
b^\kl b^\mn + \half b^\km b^\ln  - \half b^\kn b^\lm)  
\nonumber\\
&&
- \frac 32 \xi_1 (
 \et^{\ka\mu} b^\la_{\pt{\la}\al} b^{\nu\al} 
-\et^{\la\mu} b^\ka_{\pt{\ka}\al} b^{\nu\al} 
\nonumber\\
&&
\hskip 20pt
-\et^{\ka\nu} b^\la_{\pt{\la}\al} b^{\mu\al} 
+\et^{\la\nu} b^\ka_{\pt{\la}\al} b^{\mu\al} )
\nonumber\\
&&
+\frac 12 \xi_1 (\et^{\ka\mu} \et^{\la\nu} 
- \et^{\la\mu} \et^{\ka\nu}) b^{\al\be} b_{\al\be},
\nonumber\\
\ub_B &=& (\frac 32 \xi_1 - \xi_3) b^{\al\be} b_{\al\be}.
\label{stu}
\eea
Comparison of these vacuum-value coefficients 
with the results \rf{stua} 
for the coefficient fields 
appearing in the Lagrange density \rf{act2}
reveals a rescaling of the latter
of the type described in Ref.\ \cite{qbak}.

It is instructive to compare the present results 
for the antisymmetric 2-tensor
to the equivalent ones for bumblebee theories.
In these models,
a potential for a vector field $B_\mu$ 
drives the formation of a vacuum value $b_\mu$
and thereby triggers spontaneous Lorentz violation.
Possible nonminimal curvature couplings 
include Lorentz-violating couplings 
of the $s^\mn$ and $u$ types,
but $t^\klmn$ couplings cannot appear
\cite{akgrav}.
In contrast,
the theory \rf{act2} investigated here
provides an explicit example
of how nonzero $t^\klmn$ couplings can arise.
Although the coefficients $(\tb_B)^\klmn$ 
produce no leading-order contribution
to the linearized gravitational field equations,
they may generate nonzero contributions
at higher orders.
Moreover,
the coefficients $(\tb_B)^\klmn$
contain information about $b_\mn$
that is absent in $(\sb_B)^\mn$,
as can be verified by inspection 
of Eq.\ \rf{stu} in the special frame \rf{special}.
Establishing the phenomenological role 
of the coefficients $(\tb_B)^\klmn$
is an interesting open issue 
for future investigation.

Given the linearized gravitational field equations
in the form \rf{genform}
and the explicit expressions \rf{stu}
for the coefficients for Lorentz violation,
we can extract the post-newtonian metric
sourced by a given distribution of matter.
For this purpose,
we assume the matter 
is described as a conventional perfect fluid
generating the gravitational potentials
$U$, $U^{jk}$, $V^j$,
$W^j$, $X^{jkl}$, $Y^{jkl}$
defined in Eq.\ (28) of Ref.\ \cite{qbak}.
We work at post-newtonian order $O(3)$,
and choose the post-newtonian gauge at this order as 
\bea
\prt_j g_{0j} = 
\half \prt_0 g_{jj},
\quad
\prt_j g_{jk} =
 \half \prt_k
(g_{jj} - g_{00}).
\label{gauge2}
\eea
Including terms to post-newtonian order $O(3)$,
we obtain
\bea
g_{00} &=& -1 + 2U + 3 (\sb_B)^{00} U
+(\sb_B)^{jk} U^{jk} 
\nonumber\\
&&
- 4 (\sb_B)^{0j} V^j + O(4),
\nonumber \\
g_{0j} &=& -(\sb_B)^{0j}U - (\sb_B)^{k0} U^{jk}
\nonumber\\
&&
- \frac 72 (1 + \frac {1}{28} (\sb_B)^{00})V^j                 
- \frac 12 (1+\frac {15}{4} (\sb_B)^{00})W^j        
\nonumber\\
&&
+\frac 34 (\sb_B)^{jk} V^k  
+\frac 54 (\sb_B)^{jk} W^k     
+\frac 94 (\sb_B)^{kl} X^{klj}
\nonumber\\
&&
-\frac {15}{8} (\sb_B)^{kl} X^{jkl}
-\frac 38 (\sb_B)^{kl} Y^{klj},
\nonumber \\
g_{jk} &=& \de^{jk} + [(2 - (\sb_B)^{00}) \de^{jk}] U
\nonumber\\
&&
+ [ (\sb_B)^{lm} \de^{jk} - (\sb_B)^{lj} \de^{mk}
\nonumber\\
&&
-(\sb_B)^{lk} \de^{mj} + 2(\sb_B)^{00} \de^{jl} \de^{km} ] U^{lm}.
\label{pnm}
\eea
Note that the corresponding
explicit post-newtonian solutions for $\Btw_\mn$ 
can also be obtained
from the equations of motion \rf{linearb2}.

The above result for the post-newtonian metric 
involves the vacuum coefficients \rf{stu}.
However,
with the assumption of a conventional perfect fluid
and the gauge choice \rf{gauge2},
the result \rf{pnm} retains the same form as the general expression 
for the pure-gravity sector of the minimal SME.
As a consequence,
the implications for experimental and observational tests
derived in Ref.\ \cite{qbak}
apply directly to the theory \rf{act2}
in the form considered here.
For example,
the constraints on the SME coefficients $\sb^\mn$
obtained via lunar laser ranging
\cite{bcs}
and from atom interferometry
\cite{mchcc}
can be reinterpreted as limits
on $(\sb_B)^\mn$.
Other potential methods to measure these coefficients
include laboratory experiments with torsion pendula
or gravimeters,
observations of the precession of orbiting gyroscopes,
analyses of timing signals from binary pulsars,
solar-system tests involving perihelion precessions,
and time-delay and Doppler measurements
\cite{qbak,qb}. 

We conclude this section with a brief discussion
of the relation of the post-newtonian metric \rf{pnm}
to the parametrized post-newtonian (PPN) formalism
\cite{cmw,ppn}
developed for testing gravitational physics.
The PPN formalism assumes the existence 
of a special frame
in which all unconventional effects
are controlled by isotropic parameters,
so any putative match to Eq.\ \rf{pnm}
requires identifying a frame in which the coefficients
$(\sb_B)^\mn$ and $(\tb_B)^\klmn$ are isotropic.
For such a frame to exist,
the coefficients $(\sb_B)^\mn$, $(\tb_B)^\klmn$ 
must satisfy the isotropic constraints 
\bea
(\sb_B)^{0j} &=& 0,
\nonumber\\
(\sb_B)^{jk} &=& \frac 13 \de^{jk} (\sb_B)^{00},
\nonumber\\
(\tb_B)^{\ka\la\mu\nu} &=& 0.
\label{isotropic} 
\eea
However,
no such frame exists when $b_\mn$ is nonzero.
One way to see this is to use the separation \rf{vecs}
into two spatial vectors $\vec e$, $\vec b$
to write the isotropic constraints on $(\sb_B)^\mn$
in the form
\bea
\vec e \times \vec b &=& 0,
\nonumber\\
({\vec e}^2 +{\vec b}^2) \de^{jk} - 3 e^j e^k - 3 b^j b^k &=& 0.
\label{c2}
\eea
Some manipulation then reveals that only 
$\vec e=\vec b=0$ can satisfy these constraints.
The present theory for an antisymmetric 2-tensor $B_\mn$ 
with nonzero vacuum value $b_\mn$
therefore lacks an isotropic post-newtonian limit,
and hence it lies outside the PPN.
This implies no experimental or observational limits on the theory
can be placed from post-newtonian tests 
analyzed via the PPN formalism.

\section{Tadpoles in the minimal model}
\label{RGanalysis}

In this section,
we return to the minimal theory \rf{minimal} 
in Minkowski spacetime
and investigate one aspect of its quantum behavior.
While the renormalizability 
of various sectors of the SME viewed as an effective field theory 
has been studied at one loop
\cite{renorm},
less is known about the issue of renormalizability
and its relation to the potential $V$
in theories with spontaneous Lorentz breaking.
Here,
we consider the effective action
at linear order in the bare couplings
and study the behavior of the resulting interactions
under the renormalization group (RG).
The Wilson formulation of the RG
\cite{wk,wh,ahph}
has been used to adduce evidence 
for relevant nonpolynomial interactions in scalar field theories
\cite{hh1,hh2},
while exact RG methods
\cite{jp}
imply an essentially regularization-independent 
differential equation governing the RG flow for these interactions
\cite{vp}.
Similar methods can be applied to models with Lorentz violation
\cite{ba},
including bumblebee theories 
\cite{baak}.
In what follows,
we briefly summarize the scalar and vector cases
and outline results for the minimal model \rf{minimal}
involving the antisymmetric 2-tensor field $B_\mn$.
Details of the methodology 
and a summary of possible issues
can be found in Ref.\ \cite{baak}.

\subsection{Scalar and vector}

Consider a theory with a single real scalar field,
with euclidean action in $d$ dimensions 
given in terms of bare fields by
\beq
S_b= \int d^d x
(\half \prt_\mu\ph \prt^\mu\ph + V_b).
\eeq
In what follows,
the interaction Lagrange density $V_b$
is taken to be representable as a power series in $\ph^2$,
and a momentum cutoff $\La$ is used to regulate loop integrals.

Loop corrections generate the renormalized effective action $S$.
Finding an exact expression for $S$ requires
determining the coefficient of every effective $n$-particle vertex, 
which requires all $n$-point correlation functions.
This is a challenging task.
One approach yielding an approximate expression
is to limit attention to interactions 
that are at most linear in the bare couplings. 
All contributions to the effective $n$-particle amplitude 
then involve bare $(n+2k)$-point vertices 
attached via $k$ tadpole loops.
These diagrams can be summed 
\cite{keha}.
Each loop contributes a factor of $\Delta_{F}(0)/2$. 
The factor of 2 is the symmetry factor for the loop, 
while $\Delta_{F}(x-y)$ is the Feynman propagator 
for a massless scalar,
which differs from the negative inverse Laplacian 
only through its large-momentum regulation. 
In four dimensions,
$\Delta_{F}(0)=\La^{2}/16\pi^{2}$.
Each diagram also acquires an additional symmetry factor of $k!$ 
corresponding to the interchange of $k$ loops. 

In terms of dimensionless effective coupling constants,
the effective interaction action can be written as
\beq
S_b^{\rm int}=\int d^d x~ \La^{d}U_b,
\eeq
where the dimensionless potential $U_b = U_b(\La^{-(d-2)/2}\ph)$ 
depends upon $\La$ as a parameter as well as on $\ph$.
This dependence determines the nontrivial RG flow.
Including all the first-order contributions, 
$U_b$ must satisfy
\bea
&&
\hskip -50pt
\La\fr{\prt U_b}{\prt\La}+dU_b
-\half (d-2) \La^{-(d-2)/2}\ph U_b'
\nonumber\\
&&
\hskip 70pt
= -\half (d-2) C_b U_b''.
\label{rgeq}
\eea
The right-hand side of this equation 
contains the quantum corrections
and arises entirely from tadpole contributions.
The numerical value of the constant $C_b$ 
depends on the regulator,
but the result \rf{rgeq} is otherwise regulator independent
in the tadpole approximation. 
Using a cutoff regulator yields $C_b = 1/16 \pi^2$
in four dimensions.

The solutions $U_b$ of the differential equation \rf{rgeq}
with power-law dependences on $\La$,
$\La \prt U_b/\prt \La
=-\la U_b$,
are eigenmodes of the RG flow near the gaussian fixed point.
Solutions with positive anomalous dimension $\la$ 
correspond to asymptotically free theories. 
They have stronger scale dependences 
than superficially renormalizable theories
and involve relevant nonpolynomial interactions.
Each value of $\la$
gives only one functional form for the interaction,
at least at the lowest nontrivial order in the coupling. 
Each model is therefore renormalizable at this level,
being specified completely
by the value of the coupling at a fixed energy
and by the anomalous dimension $\la$,
which controls the energy dependence of the cross section.

In the above,
renormalizability is understood 
to be the statement that  
all divergences can be eliminated
and all experimental properties determined
by specifying only a finite number of observable quantities.
It may seem counterintuitive that a nonpolynomial theory 
can be renormalizable in this sense
because expanding in monomials
produces an infinite number of coefficients 
that could be deemed adjustable. 
However, 
a sum of monomials is only one way to express a function. 
For example,
although $g\exp(c\phi^{2}/\La^{2})$
can be expanded as an infinite number of monomial operators, 
the polynomial 
${\mu^{2}}\phi^{2}+{\la}\phi^{4}$ 
could also require an infinite sum
to represent it in terms of other operators.
The sine-Gordon theory in 1+1 dimensions 
has a potential with an infinite number of monomial terms, 
but the theory is known to be renormalizable
\cite{cmdhn}.
Furthermore,
the RG relevance or irrelevance 
of nonpolynomial potentials for $\ph$ 
is distinct from the known irrelevance of all monomial potentials
of degree greater than four.
The monomials fail to span the infinite-dimensional vector space 
of entire functions
and hence form an incomplete basis 
for the space of allowed potentials,
so the generic behavior of nonpolynomial theories 
cannot be inferred from the triviality 
of interacting polynomial theories.
The widespread use of the incomplete basis of monomials 
in perturbative calculations
originates in their special and convenient relationship 
to external states of known particle number,
but this feature is inessential in the RG context.

Next,
we summarize briefly
the case of an action for a vector field $B_\mu$
with Maxwell kinetic term and potential $V_b$
expressible as a power series in $B_\mu B^\mu$
\cite{ks}.
Note that this bumblebee theory has no gauge invariance.
The RG calculations in euclidean space
parallel those for a multiplet of four scalars
except for minor changes 
arising from the structure of the kinetic term
\cite{baak}.
As in the scalar case, 
the eigenmodes of the RG flow
with positive anomalous dimension
correspond to asymptotically free theories.

All nontrival potentials of this type 
generate a vacuum value $b_\mu$ 
for the bumblebee field $B_\mu$
and trigger spontaneous Lorentz breaking.
However,
only a subset lead to stable theories
in Minkowski spacetime,
where $B_\mu B^\mu$ can be either positive or negative.
Stable renormalizable theories arise 
when the anomalous dimension $\la$ is less than two 
and $b_\mu$ is spacelike
or when the anomalous dimension $\la$ is greater than eight 
and $b_\mu$ is timelike.

\subsection{Antisymmetric 2-tensor}

For the case of the antisymmetric 2-tensor $B_\mn$,
we consider the minimal theory
with Lagrange density \rf{minimal}.
Only two independent observer scalars 
can be constructed from $B_\mn$,
which we choose as 
\beq
X = B_\mn B^\mn,
\quad
Y = B_\mn \cB^\mn.
\label{xydef}
\eeq
The scalars $X_1$, $X_2$ defined in  Eq.\ \rf{x1x2}
could also be adopted,
but the above choice simplifies the presentation of the RG equation.
The bare potential $V_b$ is taken 
to be expressible as a power series in $X$ and $Y$.
As occurs for the bumblebee theory, 
the equations for the RG flow are equivalent 
up to numerical factors
to those for a multicomponent scalar field.
The differences are encoded in a fraction $f$,
which is the number of propagating degrees of freedom 
divided by the total number of degrees of freedom 
appearing in the kinetic term.
For the antisymmetric 2-tensor,
$f = 1/2$ because three of the six degrees of freedom 
have propagators in the kinetic term.

Restricting attention to interactions 
that are at most linear in the bare couplings
implies as before that the diagrams for the effective amplitudes
involve tadpole loops attached to bare vertices.
Each tadpole loop arises from a contraction 
of two factors of $B_\mn$
closing an external line,
with the specific contraction determining 
the resulting contribution from the diagram.
In forming the loops,
an external line can be closed 
in one of five ways.
Connecting the two fields within a single $X$ 
yields a factor of $12C\La^{2}$, 
where $C=2fC_b$.
For two tensors from different $X$ terms,
there are four possibilities 
generating a net contribution of $8C\La^2 X$.
Contracting two fields in a single $Y$ gives zero. 
The four ways to connect a field in $X$ with one in $Y$
yield $8C\La^2 Y$, 
while the four possibilities using tensors 
from two different $Y$ factors yield $8C\La^2 X$.
Note that there is no mixing of parity-odd and parity-even parts 
of the interaction.

To investigate the RG flow,
we write the effective potential 
in terms of dimensionless couplings $g_{j,k}$ as
\beq
\label{V}
V(X,Y)=\sum_{j,k=0}^{\infty} g_{j,k} \La^{4}
\fr{X^j Y^k}{\La^{2(j+k)}}.
\eeq
Operating on $V$ to obtain $\La dV/d\La$ 
yields two kinds of contributions, 
those derived from direct differentiation of Eq.\ \rf{V}
and those arising via $g_{j,k}$ 
from the differentiation of loop diagrams 
\cite{hh2,wh}.
Keeping only quantum corrections linear in the field,
which correspond to the tadpole diagrams,
the result is 
\begin{widetext}
\bea
\La\fr{dV}{d\La} &=& 
\sum_{j,k=0}^{\infty}
\Bigg\{
\Big[
\La
\fr{dg_{j,k}}{d\La}+4g_{j,k}-2(j+k)g_{j,k}
\Big]
\La^{4} \fr{X^j Y^k}{\La^{2(j+k)}}
\nonumber\\
&& 
\hskip 20pt
+g_{j,k}\La^{4}\fr{1}{\La^{2(j+k)}}
\Big[
12j X^{j-1} Y^k
+4j(j-1) X^{j-1} Y^k
+4jk X^{j-1} Y^k + 4k(k+1) X^{j+1} Y^{k-2}
\Big]C\La^{2}
\Bigg\}.
\qquad
\label{Vderivative}
\eea
\end{widetext}
Some combinatorial factors appear in the quantum corrections,
which are the terms involving $C\La^{2}$.

The effective potential should be independent of the cutoff, 
$\La {dV}/{d\La}=0$.
Also,
if the potential is an eigenmode of the RG flow 
near the gaussian fixed point, 
then the couplings $g_{j,k}$ 
should have power-law scaling with $\La$,
\beq
\La\fr{dg_{j,k}}{d\La}=-\la g_{j,k}.
\eeq
Here,
$\la$ is the anomalous dimension of the potential. 
Inserting these conditions into Eq.\ \rf{Vderivative} 
and equating powers of $X$ and $Y$ 
yields the recurrence relation for the couplings $g_{j,k}$ 
of an eigenmode as 
\bea
[\la-4+2(j+k)]g_{j,k}
&=&
C\big\{
4(k+1)(k+2)g_{j-1,k+2}
\nonumber\\
&&
\hskip -60pt
+\left[12(j+1)+4j(j+1)+8k(j+1)\right]g_{j+1,k}
\big\}.
\nonumber\\
\label{recurrence}
\eea
By definition, $g_{-1,k}=0$.

The recurrence relation \rf{recurrence} is more complicated 
than the equivalent expressions for the scalar and vector cases 
because three couplings are involved rather than two. 
The number of different interaction terms 
involving $2n$ powers of $B_\mn$ is $n+1$. 
If all the couplings at order $(2n-2)$ are known, 
then the couplings at order $2n$ are constrained by $n$ equations,
one for each lower-order coupling. 
This means one coupling is undetermined at each order.
For example, 
arbitrary values for the entire set $\{g_{0,k}\}$ 
can be chosen,
whereupon all other couplings are fixed.
The freedom to adjust infinitely many nonzero parameters 
is an indication of possible nonrenormalizability,
since an infinite number of measurements
is then required to specify the theory.
However,
renormalizability can be restored 
if at most finitely many parameters are nonzero.
For example,
if the effective potential depends only on $X$,
so that all the couplings $g_{0,k}$ vanish,
then a stable theory can be specified 
by the anomalous dimension $\la$ and a single coupling $g$. 
This suffices for renormalizability,
since only two measurements can fix $\la$ and $g$.

The general case has nontrivial dependence on both $X$ and $Y$.
The key feature of the theory responsible for
the possible nonrenormalizability 
is the existence of more than one independent observer scalar,
as in Eq.\ \rf{xydef}.
We therefore expect that other theories 
with general interactions involving tensors of higher rank 
also exhibit possible nonrenormalizability.
Note,
however,
that nonrenormalizable interactions may nonetheless be relevant,
since for a stable theory 
a positive anomalous dimension $\la$  
implies the effective potential grows at large scales, 
the free-field fixed point is ultraviolet stable, 
and the theory displays asymptotic freedom. 

The recurrence relation \rf{recurrence} is equivalent 
to a partial differential equation
for the effective potential.
It is convenient to introduce 
the dimensionless independent variables $x$, $y$
and dimensionless effective potential $U$ by 
\beq
x = \fr X {\La^{2}}, 
\quad
y = \fr Y {\La^{2}},
\quad
U(x,y) = \fr {V(X,Y)} {\La^{4}}.
\eeq 
Then,
the recurrence relation becomes
\bea
&
\hskip -1 truein
(\la-4)U+2xU_{x}+2yU_{y}-12CU_{x}
\nonumber\\
&
\hskip 0.5 truein
-4Cx\left(U_{xx}+U_{yy}\right)-8CyU_{xy}=0,
\label{UDE}
\eea
where partial derivatives of $U$ 
are denoted by subscripts,
$U_x$, $U_y$, etc.

We know of no exact solutions 
to Eq.\ \rf{UDE} that are both absolutely stable 
and have a nontrivial dependence on $y$.
An example of a solution with weak instability 
is the effective potential
\beq
U(x,y)=g\exp\left(\fr{x}{4C}\right)\cosh\left(\fr{y}{4C}\right),
\eeq
which satisfies Eq.\ \rf{UDE}
with anomalous dimension $\la=7$.
This potential is strictly positive 
but tends to zero as $x\rightarrow-\infty$
for $|y|<|x|$, 
so there is no lowest-energy field configuration. 
The instability is weak
because the energy approaches a limiting constant
instead of diverging negatively.
This implies tiny modifications of the potential
suffice to restore stability.
Adding a superficially renormalizable interaction
such as $g'X^2$ suffices
to obtain a physically meaningful model 
at the level of effective field theory with a finite cutoff,
and it triggers a Lorentz-violating vacuum expectation value 
with $y=0$ and large negative $x$. 
However,
the RG flow suggests the extra term is irrelevant
and fails to produce a stable continuum limit. 
It is also conceivable that stability could be restored
at the nonlinear level.

The relationship between the effective potential
and Lorentz violation is encoded in the recurrence relation
\rf{recurrence}.
If $B_\mn$ develops a nonzero vacuum value,
the theory must either spontaneously break Lorentz symmetry
or be unstable.
A necessary condition for Lorentz invariance 
is the existence of a local minimum 
of the effective potential $U$ 
at $B_\mn=0$ or,
equivalently, 
at $x=y=0$. 
This implies that
\bea
U_{x}(0,0) & = & 0 ,
\quad
U_{y}(0,0) = 0 ,
\nonumber\\
U_{xx}(0,0) & \geq & 0 ,
\quad
U_{yy}(0,0) \geq 0.
\label{uconds}
\eea
We next examine the implications of these conditions
and the recurrence relation \rf{recurrence}
at each order in $B_\mn$.

Consider first the lowest-order couplings,
corresponding to terms up to fourth order in $B_\mn$. 
The conditions \rf{uconds} imply 
\beq
g_{1,0}=g_{0,1}=0,
\quad
g_{2,0}\geq 0, 
\quad
g_{0,2}\geq 0.
\eeq
The recurrence relation \rf{recurrence}
imposes two additional linear equations 
relating the quadratic and quartic couplings,
\bea
(\la-2)g_{1,0} & = & 32Cg_{2,0}+8Cg_{0,2},
\nonumber \\
(\la-2)g_{0,1} & = & 20Cg_{1,1}.
\eea
The only way to satisfy all these conditions 
is to have $g_{2,0}=g_{1,1}=g_{0,2} =0$.
A similar argument holds at sixth order. 
Since $U(x,0)$ is required to be stationary at $x=0$, 
the coupling $g_{3,0}$ must vanish.
Likewise,
$g_{0,3}=0$. 
The recurrence relation \rf{recurrence}
then implies the remaining coefficients 
$g_{2,1}$, $g_{1,2}$
vanish at this order as well.
These analytic arguments become more subtle 
at eighth order.
Although Eq.\ \rf{recurrence} 
forces the condition $g_{3,1}=g_{1,3}=0$, 
nonzero values of the other coefficients are allowed
because the recurrence relation is satisfied
for $g_{4,0}=g_{0,4}/16$, $g_{2,2}=-3g_{0,4}/4$. 
However, 
inspection of the graph of
$(x^4/16) - (3x^2 y^2/4) + y^{4}$
reveals that $x=y=0$ is a saddle point
instead of a local extremum, 
so all the eighth-order coefficients must vanish too. 
Elementary analytic arguments of this type
suffice to show that all coefficients vanish
up to the fourteenth order in $B_\mn$.
We expect this result to hold at all orders.
Even if this conjecture is incorrect,
the above arguments show that most effective potentials
either trigger spontaneous Lorentz violation
or are unstable.

Consider now the special case of potential $V=V(B_\mn)$
depending only on the parity-even observer scalar $X$ or,
equivalently,
only on $X_1$ as defined in Eq.\ \rf{x1x2}.
This restriction implies a unique solution
to the recurrence relation \rf{recurrence}
up to an overall constant.
We find
\beq
V(B_\mn)=g\La^{4}[M(\half\la - 2,3,z) - 1],
\label{Bmnsoln}
\eeq
where the argument $z$ is given by
\beq
z=\fr X {2C\La^2} = 
\fr {8\pi^2} {\La^2} B_\mn B^\mn.
\eeq
The function $M(\al,\be,z)$ 
is the confluent hypergeometric Kummer function,
defined as
\cite{as}
\beq
M(\al,\be,z)=
1+\fr{\al}{\be}\fr{z}{1!}+\fr{\al(\al+1)}
{\be(\be+1)}\fr{z^{2}}{2!}+\cdots.
\label{kummer}
\eeq
Plots of the function $M(\al,3,z)$ can be found  
in Ref.\ \cite{jel}. 

The effective potential \rf{Bmnsoln} for $B_\mn$
is closely related to that of the effective potential $V(B_\mu)$
in the bumblebee theory.
The latter takes the form 
\cite{baak}
\beq
V(B_\mu)=g\La^{4}[M(\half\la -2,2,z) - 1],
\eeq
with $z=-32\pi^2B_\mu B^\mu/3\La^2$.
The functional properties of $V(B_\mn)$ 
and $V(B_\mu)$ are therefore similar.
In both cases,
Lorentz violation is ubiquitous.
Stable theories exist
for a range of positive values of the anomalous dimension $\la$,
and all the corresponding potentials 
exhibit spontaneous Lorentz breaking.

The effective potential \rf{Bmnsoln} allows stable theories 
with both positive and negative vacuum values $x_1$ for $X$.
Decomposing $B_\mn$ as 
$B_{0j}=-\Si^j$, 
$B_{jk} = \ep_{jkl} \Xi^l$
in analogy to Eq.\ \rf{timespacedecomp}
yields $X = \vec\Xi^2 - \vec\Si^2$,
which can be either positive or negative.
The argument $z$ 
in the effective potential \rf{Bmnsoln} 
can therefore acquire either sign
in the local minimum.
An analysis paralleling that in Ref.\ \cite{baak}
reveals that 
stable theories exist for positive $x_1$ 
when the anomalous dimension lies between zero and two.
For negative $x_1$,
stability appears for $\la$ greater than ten.
In the latter case,
metastable vacua also occur
with larger vacuum values for $X$.

\section{Summary}
\label{Summary}

In this work,
we have studied field theories with spontaneous Lorentz violation
involving an antisymmetric 2-tensor $B_\mn$.
The theories are defined through
a general class of actions of the form \rf{totact}.
The core of the action includes kinetic terms for $B_\mn$
and a potential $V$ driving spontaneous Lorentz violation.
Other components include a gravity sector,
a matter sector,
and nonminimal gravitational couplings.

All nonminimal nonderivative gravitational couplings to $B_\mn$ 
that are linear in the curvature
are displayed in Sec.\ \ref{Kinetic term}.
Sec.\ \ref{Potential term} discusses aspects of the potential,
which can be taken as a function of 
the two observer-scalar field operators
$X_1$ and $X_2$ defined in Eq.\ \rf{x1x2}.
The Lorentz-violating solutions to the equations of motion
are classified by two vacuum values,
$x_1$ and $x_2$.
Generic features of these theories
include the appearance of massless NG modes,
which are solutions of Eq.\ \rf{ngcond},
and the massive modes,
which can be identified with $X_1$ and $X_2$.
In some models,
certain NG modes appear as physical modes,
called phon modes,
that propagate at long range.
 
A comparatively simple class of theories
with some elegant features
consists of Lagrange densities 
with gauge-invariant kinetic term for $B_\mn$
and without nonminimal couplings.
These minimal models are the subject of Sec.\ \ref{Minimal model}.
We show they are equivalent to certain field theories 
with spontaneous Lorentz violation
involving a vector $A_\mu$.
In Minkowski spacetime and in the absence of Lorentz violation,
these equivalences reduce to the known dualities
between massless $B_\mn$ and scalar fields
and between massive $B_\mn$ and vector fields
\cite{op}. 
The potential for Lorentz violation
produces a hybrid duality 
in which phon mode and massive modes 
appear as different combinations of the components 
of the vector $A_\mu$.
Couplings to external currents 
and to gravity in Riemann spacetime
leave unaffected this basic picture,
as shown in Sec.\ \ref{Currents and curvature}.

Some features of nonminimal curvature couplings of $B_\mn$
are considered in Sec.\ \ref{Nonminimal model}.
In gravitational theories with spontaneous Lorentz breaking,
the dominant curvature couplings generating Lorentz violation
involve one or more of the three coefficient fields
$s^\mn$, $t^\klmn$, and $u$
\cite{akgrav}.
The action \rf{totact} for $B_\mn$
incorporates all three types of couplings.
We demonstrate this using 
the Lagrange density \rf{act2},
which is a restriction of the theory \rf{totact} 
both simple enough for illustrative purposes
and sufficiently general
to exhibit nonzero coefficient fields $s^\mn$, $t^\klmn$, and $u$.
In Sec.\ \ref{linearization},
this theory is linearized about an asymptotically flat background.
Given suitable boundary conditions,
the massive modes become frozen at this level,
and only the phon and gravitational modes propagate. 
The post-newtonian expansion for the theory
is developed in Sec.\ \ref{pn}.
This produces the nonzero vacuum values \rf{stu}
for all three coefficient fields,
a feature absent from other gravitationally coupled models
with Lorentz violation
discussed in the literature.
The post-newtonian metric is
constructed as Eq.\ \rf{pnm}.
It predicts a variety of signals
in post-newtonian tests of gravity.
Many can be measured in existing or planned searches,
while none are accessible to analyses using the PPN formalism.

In Sec.\ \ref{RGanalysis},
we return to the minimal model in Minkowski spacetime
and study the quantum behavior 
of the Lorentz-violating potential.
The RG flow in the tadpole approximation
is determined by Eq.\ \rf{Vderivative}.
An analytic solution for the special case 
with potential depending only on 
the parity-even observer scalar
is obtained in Eq.\ \rf{Bmnsoln}.
For potentials of this form,
stable theories exist
with anomalous dimensions lying between zero and two
or larger than ten.
All potentials of this type exhibit spontaneous Lorentz breaking. 

In conclusion,
the spontaneous breaking of Lorentz symmetry
via an antisymmetric 2-tensor
offers some intriguing features.
While these field theories display the properties expected 
from the broad existing treatment for general tensor fields
\cite{akgrav,bk,bfk},
the structure of the NG and massive modes
and of the gravitational couplings
arising from the case of the antisymmetric 2-tensor
implies distinctive physical content.
The properties discussed in the present work
suggest interesting possibilities
for phenomenological applications,
with definite signals that can be sought 
in present or forthcoming 
experimental and observational tests of Lorentz symmetry.

\section*{Acknowledgments}

This work was supported in part
by the United States Department of Energy
under grant DE-FG02-91ER40661.

\end{document}